\documentclass[aps,prd,showpacs,floatfix,twocolumn,
superscriptaddress,
nofootinbib,preprintnumbers,
10pt]{revtex4-2}
\makeatletter
\def\p@subsection{}
\makeatother

\usepackage{graphicx,amsmath,amssymb,lipsum,bm}

\usepackage[dvipsnames]{xcolor}
\definecolor{xlinkcolor}{rgb}{0.7752941176470588, 0.22078431372549023, 0.2262745098039215}
\definecolor{BrickRed}{rgb}{0.7752941176470588, 0.22078431372549023, 0.2262745098039215}
\definecolor{xlinkcolor}{HTML}{1c1e94}
\usepackage{booktabs} 
\usepackage[normalem]{ulem}
\usepackage[utf8]{inputenc} 
\usepackage{mathtools}
\usepackage{aux_files/aas_macros}
\usepackage{xspace}

\usepackage{ulem}
\usepackage{diagbox}

\usepackage[dvipsnames]{xcolor}
\usepackage{mathrsfs}

\usepackage[colorlinks=true,citecolor=xlinkcolor,linkcolor=xlinkcolor,urlcolor=xlinkcolor, backref=false,pdfborder={0 0 0}]{hyperref}

\newcommand{\beqa}{\begin{eqnarray}}
\newcommand{\eeqa}{\end{eqnarray}}

\newcommand{\be}{\begin{equation}}
\newcommand{\ee}{\end{equation}}
\newcommand{\beq}{\begin{equation}}
\newcommand{\eeq}{\end{equation}}

\newcommand{\bseq}{\begin{subequations}}
\newcommand{\eseq}{\end{subequations}}

\renewcommand{\ln}{\mathop{\rm ln}\nolimits}

\def\ltsima{$\; \buildrel < \over \sim \;$\xspace}
\def\gtsima{$\; \buildrel > \over \sim \;$\xspace}
\def\simlt{\lower.5ex\hbox{\ltsima}}
\def\simgt{\lower.5ex\hbox{\gtsima}}

\newcommand{\Mpch}{\, h^{-1}\mathrm{Mpc}\, }
\newcommand{\hMpc}{\, h\mathrm{Mpc}^{-1}\, }

\newcommand{\Lya}{Ly-$\alpha$\xspace}

\newcommand{\apar}{\alpha_{\parallel}}
\newcommand{\aperp}{\alpha_{\perp}}

\newcommand{\alphaap}{\alpha_{\mathrm{ap}}}

\newcommand{\aiso}{\alpha_{\rm iso}}

\newcommand{\dF}{\delta_F}
\newcommand{\dL}{\delta_L}

\newcommand{\Spar}{\Sigma_\parallel}
\newcommand{\Sperp}{\Sigma_\perp}
\newcommand{\sqdeg}{\,{\rm deg}^{-2}}
\newcommand{\bPsi}{\boldsymbol{\Psi}}

\def\gsim{\raise0.3ex\hbox{$\;>$\kern-0.75em\raise-1.1ex\hbox{$\sim\;$}}}
\def\lsim{\raise0.3ex\hbox{$\;<$\kern-0.75em\raise-1.1ex\hbox{$\sim\;$}}}

\def\beqn#1{\begin{equation}\label{#1}}
\def\eeqn{\end{equation}}

\def\beqa#1{\begin{eqnarray}\label{#1}}
\def\eeqa{\end{eqnarray}}

\def\hMpc{\, h{\text{Mpc}}^{-1}\xspace}
\def\Mpch{\, h^{-1}{\text{Mpc}}}

\def\Z2{$\mathcal{Z_2}$}

\newcommand {\ignore}[1]{}

\DeclareRobustCommand{\ion}[2]{%
\relax\ifmmode
\ifx\testbx\f@series
{\mathbf{#1\,\mathsc{#2}}}\else
{\mathrm{#1\,\mathsc{#2}}}\fi
\else\textup{#1\,{\mdseries\textsc{#2}}}%
\fi}

\newcommand{\MIT}{Center for Theoretical Physics -- a Leinweber Institute, Massachusetts Institute of Technology, Cambridge, MA 02139, USA}
\newcommand{\IAIFI}{The NSF AI Institute for Artificial Intelligence and Fundamental Interactions, Cambridge, MA 02139, USA}

\newcommand{\Kavli}{Institute of Astronomy \& Kavli Institute for Cosmology, University of Cambridge, Madingley Road, Cambridge CB3 OHA, UK}

\begin{document}

\preprint{MIT-CTP/6133}

\title{Improving cosmological distance measurements at high redshift \\ using Lyman-Alpha forest BAO reconstruction}

\author{Boryana Hadzhiyska}
\email{bth26@cam.ac.uk}
\affiliation{\Kavli}
\author{Roger de Belsunce}
\email{belsunce@mit.edu}
\affiliation{\MIT}
\affiliation{\IAIFI}

\date{\today}

\begin{abstract}
The Dark Energy Spectroscopic Instrument (DESI) measures baryon acoustic oscillations (BAO) in the Lyman-$\alpha$ (\Lya) forest and provides a robust probe of the expansion history in the matter-dominated era. Yet, non-linear structure formation broadens and shifts the BAO peak. To undo these effects, we extend standard Zel'dovich reconstruction, routinely applied to galaxy surveys, to the \Lya forest. We validate our methodology on large-volume idealized \Lya forest mocks painted on top of the \textsc{AbacusSummit} $N$-body simulations, using linear control variates to suppress sample variance and thereby reduce the error bars by a factor of up to $\approx \sqrt{50}$. In fully sampled periodic boxes, reconstruction halves the radial BAO damping scale and reduces the uncertainties on the radial ($\apar$) and transverse ($\aperp$) scaling parameters by $\approx25\%$ and $\approx15\%$, respectively. It also reduces the non-linear shift from $\Delta\apar=+0.3\%$ to $+0.10\pm0.07\%$, while $\aperp$ remains unbiased. To make this approach readily applicable to observational data, we demonstrate the impact of a survey mask with a DESI-like density of $50\,\mathrm{deg}^{-2}$. For this mask, we compare two approaches to estimate the displacement field: either directly from the \Lya forest or using an external tracer with a number density of $\bar n \simeq  10^{-3}\,h^3\mathrm{Mpc}^{-3}$, such as Lyman-break galaxies and Lyman-$\alpha$ emitters---both key targets for DESI-II. Both approaches give comparable gains and reduce the uncertainties on $\apar$ and $\aperp$ by $\approx 27\%$ and $\approx 10\%$, respectively, with mild improvements for futuristic high-density surveys with $160\,\mathrm{deg}^{-2}$. 
Correspondingly, the figure of merit improves by factors of $1.66$ for fully sampled boxes, $1.65$ for a DESI-like, and $1.76$ for a high-density survey. 
We find 
slightly smaller improvements for the cross-correlation of the \Lya forest with quasars for both survey configurations.  
\end{abstract}

\maketitle

\section{Introduction}
\label{sec}

The baryon acoustic oscillation (BAO) feature provides a robust standard ruler for measuring cosmological distances. In the early Universe, acoustic waves in the coupled photon--baryon fluid propagated until the baryon-drag epoch~\cite{1970ApJ...162..815P, 1970Ap&SS...7....3S}. The distance they traveled, the sound horizon $r_d$, was imprinted as a characteristic comoving scale in the cosmic microwave background (CMB;~\cite{Planck:2018vyg}) and in the matter distribution~\cite{EisensteinHu1998,Weinberg2013}. At late times, measuring this scale across and along the line of sight constrains the transverse comoving distance and the Hubble distance, respectively, in units of $r_d$~\cite{Eisenstein2005,Cole2005, 2011MNRAS.416.3017B, 2011MNRAS.415.2892B, Alam:2021, DESI_BAO_2024}. Calibrating this standard ruler with the CMB or with big-bang nucleosynthesis~\cite{Cooke:2017cwo} turns these into absolute distance measurements.

The Stage-IV Dark Energy Spectroscopic Instrument (DESI) measures the BAO with galaxy and quasar clustering up to $z\simlt 2$~\cite{DESI2016, DESI_BAO_2024, DESI2025DR2Cosmo}. At higher redshift, DESI measures the BAO in the \Lya forest, a series of absorption features produced by intervening neutral hydrogen in the spectra of distant quasars~\cite{McQuinn:2016}. This is DESI's highest-redshift BAO measurement: from more than $820{,}000$ spectra, it determines the BAO scaling parameters at an effective redshift of $z_{\rm eff}\approx 2.33$ to 1.1\% along and 1.3\% across the line of sight~\cite{DESI2025DR2Lya}. Recently, DESI has provided evidence for evolving dark energy, driven by a discrepancy between the distances measured with DESI BAO and those predicted by the base $\Lambda$CDM model\footnote{As in the \textit{Planck} papers, we refer to the six-parameter $\Lambda$CDM model (spatially flat, power-law scalar adiabatic fluctuations, cosmological constant) as the base $\Lambda$CDM model~\cite{Planck:2018vyg}.} calibrated on the CMB~\cite{DESI2025DR2Cosmo}. This has spurred a slew of alternative explanations to reconcile the CMB and DESI distances including negative spatial curvature~\cite{Chen:2025mlf}, a higher optical depth to reionization~\cite{Sailer:2025lxj,Jhaveri:2025neg,Sullivan2026HighTau}, or models with an interacting dark sector~\cite{Khoury:2026svx}. Improving distance measurements at high redshift provides a unique lever arm on the expansion history, which can help discriminate between these models~\cite{DESI:2026lnd}.

The \Lya forest traces the matter distribution at high redshift ($2 \simlt z \simlt 4$) down to small scales, with dense sampling along individual lines of sight and sparse sampling in the transverse direction. The rapidly growing DESI \Lya sample, together with advances in three-dimensional modeling and measurement techniques~\cite{Ivanov:2023yla, deBelsunce2026FieldLevel, Belsunce_Sullivan_skewspectrum,deBelsunce:2024knf}, makes the forest an increasingly powerful cosmological observable. On large scales, it constrains the expansion history through the BAO~\cite{McDonald:2007,Slosar2013, Busca:2013, dMdB:2020,DESI_lya_2024, DESI2025DR2Lya}, and it has been used to measure the broadband shape of the large-scale three-dimensional \Lya correlation function~\citep{Slosar2013, Cuceu:2021, Cuceu:2023, Gordon:2023, Cuceu:2025nvl}. On small scales, the one-dimensional power spectrum (P1D)~\citep[see, e.g.,][]{Seljak:2005, Viel:2005,McDonald06, PYB13, Chabanier:2019, Pedersen:2020, 2023MNRAS.526.5118R, 2024MNRAS.tmp..176K} yields constraints neutrino masses~\citep{Seljak:2005, Viel:2010, PYB13, Palanque2020}, primordial black holes~\citep{Afshordi:2003, Murgia:2019,Ivanov:2025pbu}, and enables tests of (warm) dark matter models~\citep{Viel:2013, Baur:2016, Irsic17, Kobayashi:2017, Armengaud:2017, Murgia:2018,Garzilli:2019, Irsic:2020, Rogers:2022, Villasenor:2023, Irsic:2023}, early dark energy models~\citep{2023PhRvL.131t1001G}, and the thermal properties of the ionized intergalactic medium~\citep{Zaldarriaga:2002, Meiksin:2009,McQuinn:2016, Viel:2006, Walther:2019, Bolton:2008, Garzilli:2012, Gaikwad:2019, Boera:2019, Gaikwad:2021, Wilson:2022, Villasenor:2022}.

Current \Lya forest BAO analyses use linear-theory templates in which the non-linear damping of the BAO amplitude by incoherent bulk motions enters only through a Gaussian broadening, and which do not capture the out-of-phase contributions from non-linear clustering that shift the observed BAO position~\cite{EisensteinSeoWhite2007,CrocceScoccimarro2008,PadmanabhanWhite2009,SherwinZaldarriaga2012}. For the \Lya forest, these contributions shift the BAO scaling parameters at the $0.3\%$-level~\cite{deBelsunce:2024rvv} and can be modeled using the effective field theory (EFT) power spectrum modeling at one-loop\footnote{This perturbative framework has been successful in describing large-scale statistics using only the relevant symmetries, the equivalence principle, and dimensional analysis~\cite{2009JCAP...08..020M, Baumann:2010tm, Carrasco:2012cv, Ivanov:2023yla, Givans:2020sez}.}~\cite{Hadzhiyska:2025cvk}. However, marginalizing over the comparatively large set of bias (or EFT nuisance) parameters can dilute the BAO constraints in the absence of informative simulation-based priors~\cite{Ivanov:2024xgb}.

In this work, we aim to undo the damping and shift of the BAO using BAO reconstruction -- a standard technique in galaxy clustering analyses~\cite{Eisenstein2007,Padmanabhan2012,Burden2014,Seo2016,ChenVlahWhite2019,Paillas2025,Chen2024Recon}. It estimates a smoothed large-scale displacement field from the observed density, using an assumed bias and redshift-space response, and shifts the tracers and a reference sample backwards using said displacement. This sharpens the acoustic feature and suppresses the leading non-linear shift. Applying reconstruction to the \Lya forest raises two additional difficulties. First, on large scales the flux decrement responds not only to the matter density contrast, $\delta$, but also to the line-of-sight velocity gradient~\cite{McDonald:1999dt, McDonald:2001}, $\delta_F = b_F\,\delta + b_{\eta}\,\eta$ with $\eta \equiv -\frac{\partial_{\parallel} v_{\parallel}}{\mathcal{H}}$ where $v_{\parallel}$ is the line-of-sight peculiar velocity and $\mathcal{H}$ is the conformal Hubble rate. Two key differences between galaxies and the \Lya forest are that, first, the \Lya forest has a negative linear density response, $b_F<0$, and an independent velocity-gradient response, $b_\eta$, which sets its redshift-space distortion parameter, $\beta_F=f\,b_\eta/b_F$, for linear growth rate $f$. For a standard galaxy sample without line-of-sight selection effects, the corresponding coefficient is fixed to unity by the redshift-space mapping, so that $\beta=f/b$~\cite{DesjacquesJeongSchmidt2018}. Second, the \Lya forest is an almost continuous field rather than a set of discrete tracers with a dense (sparse) sampling along (across) the line-of-sight.

We address both issues in this work and apply BAO reconstruction to the \Lya forest. We test our methodology on large-volume \Lya forest mocks painted on top of the \textsc{AbacusSummit} $N$-body simulations~\cite{Hadzhiyska:2023}. 
We use linear control variates, effectively reducing the error bars 
by a factor of approximately up to $\sqrt{50}$ on the largest scales, to test for biases in the BAO scaling parameters at high precision. We consider three reconstruction regimes: In the first, we use a fully sampled periodic box, which is not applicable to real data but isolates the physics and provides a reference point for the achievable improvement. We then restrict the mocks to sightline densities representative of DESI and of a high-density DESI-II-like sample~\cite{Schlegel2022Roadmap}, and compute the displacements either from an overlapping tracer population or from the \Lya forest itself. As external tracers we use dark matter halos with $\bar n=10^{-4}$--$5.7\times10^{-3}\,h^3{\rm Mpc}^{-3}$ as proxies for \Lya emitters (LAEs) and Lyman-break galaxies (LBGs), both key observational targets of DESI-II. This allows us to assess the applicability of the method to current \Lya forest data from DESI, its successor DESI-II, and to future surveys such as the WEAVE-QSO survey~\cite{2016sf2a.conf..259P}, the Prime Focus Spectrograph (PFS;~\cite{2022PFSGE}), and 4MOST~\cite{2019Msngr.175....3D}.

The paper is organized as follows. We introduce the \textsc{AbacusSummit} \Lya forest mocks in Sec.~\ref{sec:mocks} and describe the reconstruction method in Sec.~\ref{sec:recon}. Section~\ref{sec:measure} presents the analysis procedure. In Sec.~\ref{sec:results} we present our results, and we conclude in Sec.~\ref{sec:conclusions} with a short discussion of the limitations of the present approach.

\section{Simulations}
\label{sec:mocks}

We use the \textsc{AbacusSummit} \Lya forest mocks which consist of a suite of six realizations using a fiducial \emph{Planck} 2018 best-fit cosmology~\cite{Planck:2018vyg}. These $N$-body simulations have $6912^3$ particles in periodic boxes of side length $L=2000\, \Mpch$~\cite{Maksimova2021,Hadzhiyska:2023}, and we use the outputs at a fixed redshift of $z=2.5$. The \Lya forest is painted on top of the dark matter field using a fluctuating Gunn--Peterson approximation (FGPA;~\cite{Croft:1997jf}) with added small-scale fluctuations to match the one-dimensional power spectrum from a reference hydrodynamic simulation. Briefly summarized,  the peculiar velocities map the absorption into redshift space, and the free parameters of the FGPA model are calibrated against the IllustrisTNG hydrodynamical simulation~\cite{2010MNRAS.401..791S, Weinberger:2019tbd}. This yields four \Lya forest variations, denoted by Models 1--4, with two \Lya forest realizations per box with redshift-space distortions applied either along the $\Hat{\bf{z}}$ axis or the $\Hat{\bf{y}}$ axis, denoted by LOSz and LOSy, respectively. Each Model has a different set of bias parameters, summarized for the linear bias parameters in the first two columns of Tab.~\ref{tab:response}. The main difference between the different FGPA prescriptions is that Models 3 and 4 have a larger RSD-bias parameter, $\beta_F = f\,b_\eta/b_F$ for growth rate $f$, than Models 1 and 2 with $\beta_F\simeq2.1$--$2.4$ instead of $\beta_F\simeq 1$. The full set of one-loop EFT bias parameters, i.e. the higher-order bias parameters, are given in Ref.~\cite{Hadzhiyska:2025cvk}. We refer the reader to Ref.~\cite{Hadzhiyska:2023} for a fuller presentation on how the mocks are constructed.

To mimic a DESI-like observational strategy we apply a mask in the transverse direction. Therefore, we  choose a random subset of skewers with a mean separation of $9.9\Mpch$, and a surface density of $50\sqdeg$ corresponding to $40{,}632$ sightlines per box. To mimic a high-density survey we apply a second variant with $160\sqdeg$. This denser case lies between the quasar densities planned for the next stage of spectroscopic surveys ($\simeq80\sqdeg$ \cite{SpecS5_2025}) and the densities reachable by using faint galaxies as background sources with $100-1000\sqdeg$~\cite{Lee2014Requirements}. We emphasize that these mocks are highly idealized and do not include any DESI and \Lya forest-specific contaminants, e.g.,~the instrument response, pixel noise, or continuum fitting. 

\section{BAO reconstruction}
\label{sec:recon}

We use the transmitted flux fluctuation, $\dF=F/\overline{F}(z)-1$, around the mean transmission $\overline{F}(z)$ at redshift $z$. On large scales, the forest is described by the well-known Kaiser-like formula~\cite{McDonald:1999dt,McDonald:2001}
\begin{equation}
  \dF(\mathbf k)=b_F\left(1+\beta_F\mu^2\right)\dL(\mathbf k),
  \label{eq:response}
\end{equation}
where $\dL$ is the linear matter density, $\mu$ the cosine between $\mathbf k$ and the line of sight, $b_F<0$ the linear density bias, and $\beta_F=f\,b_\eta/b_F$ the redshift-space distortion parameter set by the velocity-gradient bias, $b_\eta$, and the growth rate, $f$. For each model, we measure $b_F$ and $\beta_F$ from the cross-spectrum of the \Lya forest with the phase-matched initial density field over $0.005\leq k<0.05\hMpc$, given in Tab.~\ref{tab:response}. For observational data, we would obtain these from fits to the large-scale \Lya forest auto-correlation function.

We estimate the Zel'dovich displacement, $\bPsi$, by solving the linear redshift-space equation~\cite{Eisenstein2007,Padmanabhan2012,Seo2016}
\begin{equation}
  \nabla\cdot\bPsi+\beta_F\,\nabla\cdot\left[(\bPsi\cdot\hat{\mathbf z})
  \hat{\mathbf z}\right]=-\frac{S\star\dF}{b_F},
  \label{eq:solve}
\end{equation}
where $\hat{\mathbf z}$ is the line of sight (in the plane-parallel approximation) and $S\star\dF$ is the flux field smoothed with the Gaussian kernel $S(k)=\exp(-k^2R^2/2)$, with $R=10\Mpch$.\footnote{We  investigate  the impact of varying the smoothing scale from $4$ to $20\Mpch$ for Model~1 for one realization (phase one) in App.~\ref{fig:smoothing} and find that the radial gain has a broad maximum at $R=8$--$12.5\Mpch$ and falls to $1.36$ at $R=20\Mpch$ with the transverse gain slowly decreasing with increasing smoothing radius $R$, from $1.23$ to $1.14$. The chosen radius is the smallest radius for which both post-reconstruction damping scales remain resolved, and the gains change by less than $3\%$ between $R=8$ and $12.5\Mpch$. } The smoothing restricts the displacement to large scales, where the Zel'dovich approximation holds and $\dF/b_F$ traces the linear density. Compared to the standard equation for galaxies, the \Lya forest has the additional velocity-gradient bias which is why we divide by $b_F$ instead of a galaxy bias with the second term carrying $\beta_F$ instead of $f/b$. Inserting Eq.~(\ref{eq:response}) into Eq.~(\ref{eq:solve}), the factor $(1+\beta_F\mu^2)$ cancels, and for a curl-free displacement the solution in Fourier space is
\begin{equation}
  \bPsi(\mathbf k)=\frac{i\mathbf k}{k^2}\,S(k)\,\dL(\mathbf k),
  \label{eq:psi}
\end{equation}
i.e.\ the real-space Zel'dovich displacement of the smoothed linear density field. 
In Tab.~\ref{tab:response} we quote the cross-correlation coefficient, $r_\Psi$, between the smoothed source estimated from the \Lya forest directly and the linear-density prediction. The estimated displacement matches the linear one with negligible error on the amplitude. Similarly, the cross-transfer function, $T_\Psi$, is unity to within $0.1\%$. 

\begin{table}[t]
\begin{ruledtabular}
\begin{tabular}{ccccc}
Model & $b_F$ & $\beta_F$ & $r_\Psi$ & $T_\Psi$\\
\hline
1 & $-0.1481$ & $0.958$ & $0.987$ & $0.999$\\
2 & $-0.1309$ & $0.990$ & $0.987$ & $0.999$\\
3 & $-0.1325$ & $2.075$ & $0.990$ & $1.000$\\
4 & $-0.1285$ & $2.395$ & $0.989$ & $1.001$\\
\end{tabular}
\end{ruledtabular}
\caption{\textbf{Linear response of the forest:} Linear bias, $b_F$, and redshift-space distortion parameter, $\beta_F$, of each model, measured from the cross-spectrum of the forest with the phase-matched initial density field over $0.005\leq k<0.05\hMpc$. The last two columns give the cross-correlation coefficient, $r_\Psi$, and the cross-transfer function, $T_\Psi$, between the smoothed displacement source estimated from the forest and its prediction from the initial density, averaged over the modes used in the $R=10\Mpch$ solve.}
\label{tab:response}
\end{table}


We solve Eq.~(\ref{eq:solve}) with the multigrid solver of \texttt{pyrecon}\footnote{Publicly available at \url{https://github.com/cosmodesi/pyrecon}.} and follow the \textsc{RecSym} convention~\cite{White:2015eaa}, in which the data and the randoms are shifted by the same redshift-space displacement, so that the linear redshift-space distortions are preserved on large scales~\cite{White:2015eaa}.\footnote{For the \Lya forest, ``randoms'' are a misnomer as they are the observed pixels with unit weights, encoding the window function.} The redshift-space displacement is given by
\begin{equation}
  \bPsi_s=\bPsi+f\,(\bPsi\cdot\hat{\mathbf z})\,\hat{\mathbf z}\,,
  \label{eq:shift}
\end{equation}
where $f$ is the linear growth rate of matter, $f \equiv \mathrm{d} \ln D/\mathrm{d} \ln{a}$, where $D$ is the linear growth factor evaluated at $z=2.5$.

We move each sample from $\mathbf x_i$ to $\mathbf y_i=\mathbf x_i-\bPsi_s(\mathbf x_i)$ and repaint the flux values with a triangular-shaped-cloud kernel, $W$, with an effective width of $2.9--5.9\hMpc$, to obtain the reconstructed flux decrement
\begin{equation}
  \dF^{\rm rec}(\mathbf y)=\frac{\sum_i\dF(\mathbf x_i)\,W(\mathbf y-\mathbf y_i)}
  {\sum_iW(\mathbf y-\mathbf y_i)}\,,
  \label{eq:paint}
\end{equation}
where the denominator encodes the survey geometry. 
At linear order, Eq.~(\ref{eq:paint}) retains the large-scale field while removing the physics responsible for the damping and shifting of the BAO. We generate a second copy of the field where we do not apply the displacements such that painting and normalization are identical pre and post reconstruction.

We consider three reconstruction regimes. 
The main difference is the field from which we solve for the displacements and how the observed \Lya forest flux pixels are moved. All three use the smoothing radius $R=10\Mpch$, solve for the displacement on the same $512^3$ mesh, and apply the shift of Eq.~(\ref{eq:shift}).

\textbf{I -- Dense self-reconstruction:} We use the fully sampled box both for estimating the displacement and reconstructing the flux decrement. For the configuration-space checks of Model~1, we also paint the moved field on a higher resolution $1152^3$ mesh with cell size $1.7\Mpch$ while the displacement is obtained from a $512^3$ mesh.\footnote{For consistency with Ref.~\cite{Hadzhiyska:2025cvk} we use a cell size of $1.7\Mpch$ which has no impact on the displacements due to the large smoothing kernel.}
This idealized case ignores the observational window function and is therefore not applicable to real data, but it gives an upper limit of the promise of our methodology.

\textbf{II -- External displacement:} We estimate $\bPsi$ from a discrete tracer at the same redshift, evaluate it at the observed \Lya forest pixels, and move these pixels. This is relevant for future surveys such as DESI-II, which are expected to provide dense samples of Lyman-break galaxies and Lyman-$\alpha$ emitters at $2<z<3.5$ that can serve as backlights for the \Lya forest~\cite{Schlegel2022Roadmap,SpecS5_2025,Ravi2024LAE,Payerne2025LBG}. The advantage of this approach is that it does not depend on the observed flux and thus on the exact values of the linear bias parameters and is not affected by \Lya-specific systematics such as continuum fitting. We use dark matter halos as proxies for high-redshift galaxies identified with \texttt{CompaSO} with at least $150$ particles ($M\ge3.2\times10^{11}h^{-1}M_\odot$; $\bar n=5.7\times10^{-3}\,h^3{\rm Mpc}^{-3}$) and a large-scale bias of $b=2.695$. We reconstruct them with the standard galaxy reconstruction procedure of Eq.~(\ref{eq:solve}), i.e.\ with $b$ in place of $b_F$ and $f/b$ in place of $\beta_F$, using \texttt{pyrecon} with periodic boundary conditions and no randoms. To emulate sparser galaxy samples, we randomly thin the halos to $\bar n=10^{-3}$ and $10^{-4}\,h^3{\rm Mpc}^{-3}$, which preserves their bias. We use the same sightlines and the same unreconstructed reference fields as in Regime~III, so that only the source of the displacement differs.

\textbf{III -- Sparse self-reconstruction:} We draw a fixed random subset of the $512^2$ transverse mesh columns as sightlines, each spanning the full box along the line of sight. We compare two number densities: First, one with $50\sqdeg$ emulating a DESI-like survey at the comoving distance of $4022\Mpch$ to $z=2.5$ corresponding to $40,632$ sightlines with a mean separation of $9.9\Mpch$. Second, a high-density survey emulating a DESI-II-like survey with $160\sqdeg$. We emphasize that only the \Lya forest pixels from the masked box are used to estimate the displacements and used to displace the field. We apply two sets of weights: for the data part we use $1+\dF/b_F$, and for the randoms we use unit weights. The smoothed source is the weighted average of the flux on nearby sightlines with the smoothing kernel interpolating between sightlines. Note that the mean separation between skewers needs to be smaller than the smoothing radius. Whilst regime III is readily applicable to observational data, it still requires validation of \Lya-specific systematics.\footnote{Conceptually this is similar to \Lya forest tomography~\cite{Horowitz2019, Horowitz2021, Ravoux2020}.}.
In Regimes~II and~III, the moved pixels keep their flux values, weights, and sightline identifiers, so that we measure them directly with the pair-count spectral estimator, introduced in Sec.~\ref{sec:sparse_est}.

\section{Analysis Method}
\label{sec:measure}

In this section, we introduce the BAO model and likelihood (Sec.~\ref{sec:model}), the linear control variates that we use to reduce the sample variance in our tests for biases (Sec.~\ref{sec:lcv}), and the measured summary statistics (Sec.~\ref{sec:sparse_est}).

\subsection{BAO model and likelihood}
\label{sec:model}

Our baseline model adopts the peak-only dilation and spline broadband of the DESI galaxy BAO analysis~\cite{Chen2024BAOTheory, Hadzhiyska:2025cvk}:
\begin{align}
  P(k,\mu)={}&b_F^2(1+\beta_F\mu^2)^2P_{\rm nw}(k)+B_w(k)\nonumber\\
  &+b_F^2(1+\beta_F\mu'^2)^2\,\frac{P_{\rm w}(k')\,D(k',\mu')}{\apar\aperp^2},
  \label{eq:model}
\end{align}
where $P_{\rm w}=P_{\rm lin}-P_{\rm nw}$ is the oscillatory part of the linear spectrum, $P_{\rm lin}$, with $P_{\rm nw}$ being its smooth component, and $B_w(k)$ is a smooth broadband term in each angular wedge $w$. The non-linear damping is captured by
\begin{equation}
  D(k,\mu)=\exp\left\{-k^2\left[\mu^2\Spar^2+(1-\mu^2)\Sperp^2\right]/2\right\}\,,
  \label{eq:damping}
\end{equation}
with two damping scales $\Spar$ and $\Sperp$, and the Alcock--Paczy\'nski (AP) mapping~\cite{AlcockPaczynski1979} is defined as 
\begin{equation}
k'_\parallel\equiv k\mu/\apar\,, \qquad k'_\perp\equiv k\sqrt{1-\mu^2}/\aperp\,,
\end{equation}
for the radial, $\apar$, and transverse, $\aperp$, BAO scaling parameters, which we also recast into~\citep{DESI_lya_2024}
\begin{equation} \label{eq:AP}
    \aiso=\apar^{1/3}\aperp^{2/3}\,, \qquad \alphaap = \apar/\aperp\,.
\end{equation}
Note that we only dilate the oscillatory part. For the templates, we use the linear power spectrum provided by \textsc{AbacusSummit} and split it into an oscillatory, $P_{\rm w}$, and smooth component, $P_{\rm nw}$, using a Savitzky--Golay filter~\cite{SavitzkyGolay1964}.\footnote{Publicly available at \url{https://github.com/cosmodesi/cosmoprimo}.} For $B_w(k)$, we use an independent cubic spline in each wedge with a knot spacing of $\Delta k = 0.06\hMpc$, following Ref.~\cite{Chen2024BAOTheory}, and  average the model in each $(k,\mu)$ bin. For our fits, the free parameters are $\apar$ and $\aperp$, $\beta_F$, and $\Spar$ and $\Sperp$. We apply broad priors to each. The amplitude, $b_F^2$, and the spline coefficients are solved analytically by linear least squares at every step. For the BAO fits we use the fitting range $0.02\leq k\leq0.40\hMpc$.\footnote{We have verified that varying $k_{\rm max}$ between $0.25$ and $0.40\hMpc$ and $k_{\rm min}$ between $0.015$ and $0.04\hMpc$ does not change the main conclusions of this work.} 
In configuration space, we fit $\xi(r_\parallel,r_\perp)$ over $30<r<180\Mpch$ with \textsc{Vega}\footnote{Publicly available at \url{https://github.com/andreicuceu/vega}.}, using the same template and peak-only dilation. We Hankel transform the spline broadband to configuration space using all multipoles up to $\ell=6$ and apply a low-pass filter, discussed in Sec.~\ref{sec:sparse_est}.

We also fit the cross-correlation between the \Lya forest and dark matter halos which we use as proxy for high-redshift quasars~\cite{Font:2013}. Its peak-only template is obtained from Eq.~\eqref{eq:model} by replacing one occurrence of the \Lya forest linear-bias factor, $K_1^F(\mathbf{k})\equiv b_F-b_\eta f \mu^2$, with the one for quasars, $K_1^Q(\mathbf{k})\equiv b_Q+ f \mu^2$, which results in the geometric mean of both tracers:
\begin{align}
P_{FQ}(k,\mu)={}&b_Fb_Q(1+\beta_F\mu^2)(1+\beta_Q\mu^2)P_{\rm nw}(k)\nonumber\\
&+B^{FQ}_w(k)+b_Fb_Q(1+\beta_F\mu'^2)\nonumber\\
&\times (1+\beta_Q\mu'^2)\frac{P_{\rm w}(k')D_{FQ}(k',\mu')}{\apar\aperp^2},
\end{align}
where $B^{FQ}_w(k)$ and $D_{FQ}(k,\mu)$ are independent cross-correlation broadband and damping terms. For the mock tests, we fix $b_Q=3.39$ and $\beta_Q=0.28$, as measured from the cross-correlation of the QSO field with the linear density field. 

We use a Gaussian likelihood throughout this work. For the power spectrum, $P(k,\mu)$, we use a Gaussian analytic covariance which is diagonal $2P^2(k,\mu)/(N_{\rm box}N_{\rm mode})$, for the measured mean spectrum  $P$, the number of expected Fourier modes per bin $N_{\rm mode}$, and the number of realizations entering the fit $N_{\rm box}$. For the multipoles measured with our spectral estimator, introduced in Sec.~\ref{sec:sparse_est} we use a Gaussian covariance computed from the pre-reconstruction multipoles. For $\xi(r_\parallel,r_\perp)$, we use the covariance matrix of the eBOSS DR16 \Lya forest auto-correlation~\cite{dMdB:2020}, divided by 20, as a fixed weight rather than as a calibrated covariance for our simulation volume. The arbitrary normalization factor follows Ref.~\cite{Hadzhiyska:2025cvk} to obtain a reduced $\chi^2$ statistic that is reasonably close to unity. We compute the error bars on $\apar$ and $\aperp$ from one-dimensional profile likelihoods such that for each proposed value of the scaling parameters, we minimize the $\chi^2$ over all other parameters. The resulting error is obtained from the $\Delta\chi^2=1$ interval. For $\aiso$ and $\alphaap$, we propagate the covariance of $\apar$ and $\aperp$, which we obtain from the second derivative of the profiled $\chi^2$ around the best fit. 
In the joint fit of the auto and cross-correlation, $\apar$ and $\aperp$ are shared between both spectra, while their amplitudes, redshift-space parameters, damping scales, and broadband terms are allowed to vary independently. Further, the Gaussian covariance includes the joint covariance~\cite[see, equation 4 in~][]{Chudaykin:2025gsh}.

To quantify the performance of our BAO reconstruction procedure applied to the \Lya forest, we use two key metrics: First, the precision gain, $G=\sigma_{\rm pre}/\sigma_{\rm post}$, which compares the errors of a single fit to the mean spectrum of the six boxes before and after reconstruction, with each result weighted by its own covariance. For the fully sampled boxes, this corresponds to a noiseless volume of $48\,h^{-3}{\rm Gpc}^3$. The precision gain quantifies the sharpening of the BAO feature. The paired response, $\Delta\alpha=\alpha_{\rm post}-\alpha_{\rm pre}$, is measured for each phase and we perform the fit with the covariance of the raw pre-reconstruction spectrum. We average the two lines of sight within each phase and quote the standard error over all six phases.

\subsection{Linear control variates technique}
\label{sec:lcv}

To test for biases in the BAO scaling parameters at high precision, we apply linear control variates (LCV;~\cite{HadzhiyskaCV2023,Hadzhiyska:2025cvk}), which subtract the part of each measured spectrum that is predictable from the known initial conditions and is a simplification of the Zel'dovich control variate technique introduced in Refs.~\cite{Kokron2022,DeRose2023,HadzhiyskaCV2023}. For each phase, we form the linear forest field, $\dF^L=b_F(1+\beta_F\mu^2)\dL$, from the phase-matched initial density with $b_F$ and $\beta_F$ of Tab.~\ref{tab:response}, and replace
\begin{equation}
  \hat P_F\to\hat P_F-c(k,\mu)\left[\hat P_L-P_L\right]\,,
  \label{eq:lcv}
\end{equation}
where $\hat P_F$ is the measured \Lya forest spectrum, $\hat P_L$ the measured power of $\dF^L$, and $P_L$ its expectation. The coefficient $c=(P_{FL}/P_{LL})^2$, with $P_{FL}$ the cross-spectrum of the \Lya forest with $\dF^L$ and $P_{LL}$ the power of $\dF^L$, is estimated from the six phases in each $\mu$ wedge, set to unity below $k=0.05\hMpc$, smoothed in $k$, and tapered around $k\simeq0.6\hMpc$, where the linear field decorrelates from the \Lya forest. Since $\hat P_L$ averages to $P_L$, Eq.~(\ref{eq:lcv}) leaves the expectation of the spectrum unchanged, effectively removing most of the linear sample variance. 
For the correlation function, we apply the correction to the Fourier modes before applying the inverse Fourier transform to the LCV-reduced 3D field, with a coefficient that depends on $k$ only and is estimated without the phase being corrected. 

\subsection{Measured two-point statistics}
\label{sec:sparse_est}

We assess the impact of the BAO reconstruction on two-point statistics. For the fully sampled boxes, we compute the power spectrum, $P(k,\mu)$, of the $512^3$ field with a grid-based fast Fourier transform (FFT) in five equal $\mu$ wedges and bins of $\Delta k=0.005\hMpc$ and correct for the particle deposition scheme. We compute the correlation function, $\xi(r_\parallel,r_\perp)$, by inverse FFT of the same power spectrum, multiplied by a smooth low-pass filter that is unity below $k=0.35\hMpc$ and vanishes at $0.5\hMpc$, and average it in bins of $4\Mpch$ in $r_\parallel$ and $r_\perp$ up to $200\Mpch$ to follow the analysis choices of DESI~\cite{DESI2025DR2Lya}.

This grid-based procedure is, however, not applicable to real data in the presence of a window function, as it introduces large aliasing effects~\cite{Font-Ribera:2018,deBelsunce:2024knf, Horowitz:2024nny, Belsunce_Sullivan_skewspectrum}. For the sparse sightlines of Regimes~II and III, we therefore use the pair-count spectral estimator \texttt{HIPSTER}~\cite{Philcox:2019hdi, Philcox:2020xod}, as adapted to the \Lya forest in Ref.~\cite{deBelsunce:2024knf}.\footnote{Publicly available at \url{https://hipster.readthedocs.io/en/latest/}.} It weights each pixel pair by $\exp(i\mathbf{k}\cdot \mathbf{r}_{ij})$, for wave vector $\mathbf{k}$ and pair separation $\mathbf{r}_{ij}$, and thus measures the anisotropic power spectrum without the need for a grid-based FFT, while accounting for the window function in a tractable way. For the present analysis we remove self-skewer contributions. We measure the power spectrum with the same pixels, flux values, weights, and sightline IDs pre and post reconstruction with the only difference that the pixel positions change. For computational efficiency, we restrict the pair separation to $R_0=200\Mpch$ with a smooth taper, $W(r;R_0)$, which we include in the theory model, and rebin each spectrum by a factor of two. We fit the monopole and quadrupole in linear bins of $\Delta k=0.0075\hMpc$ over $0.031<k<0.30\hMpc$ with three broadband terms and Gaussian priors on $\Spar$ and $\Sperp$ of width $2\Mpch$.

\section{Results}
\label{sec:results}

In this section, we present the results of BAO reconstruction on the \textsc{AbacusSummit} \Lya forest mocks. We gradually increase the level of realism and start with the idealized scenario of the fully sampled forest (Regime~I, Sec.~\ref{sec:res_recon_I}), for which we quantify the sharpening of the BAO feature, the gain in precision, and the residual shift of the BAO scale. We then restrict the forest to sparse sightlines emulating a DESI-like and a futuristic high-density \Lya forest sample. For both we compare the improvements obtained after reconstruction using the \Lya forest itself (Regime~III) and when using external tracer (Regime~II) in Sec.~\ref{sec:res_sparse}.

\subsection{Idealized BAO reconstruction}
\label{sec:res_recon_I}

\begin{figure*}
  \includegraphics[width=0.95\textwidth]{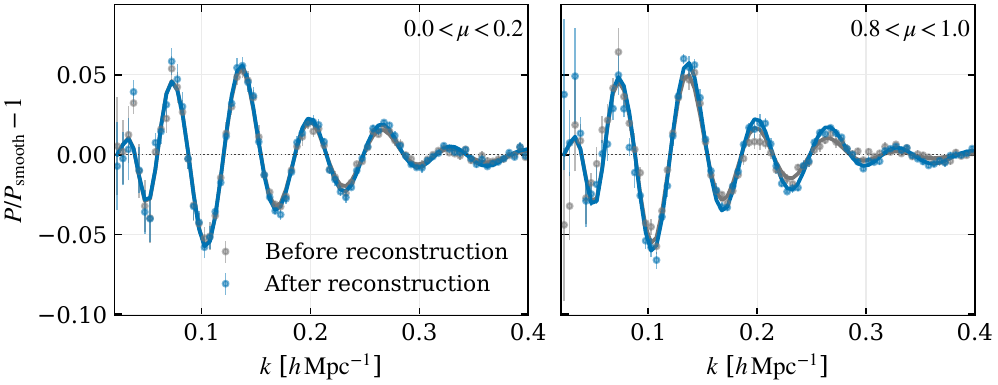}
  \caption{\textbf{Impact of BAO reconstruction in Fourier space:} BAO feature of Model~1 before (gray) and after (blue) reconstruction in the wedge closest to the transverse direction (left) and in the radial wedge (right). The points are the mean over the 12 realizations (six phases and two lines of sight), divided by the smooth part of the best fit, with error bars from the scatter between the realizations. Over $0.15<k<0.35\hMpc$, the rms of the best fit increases by $12\%$ in the transverse wedge and by $34\%$ in the radial wedge.}
  \label{fig:wiggles}
\end{figure*}

\begin{figure*}
  \includegraphics[width=0.84\textwidth]{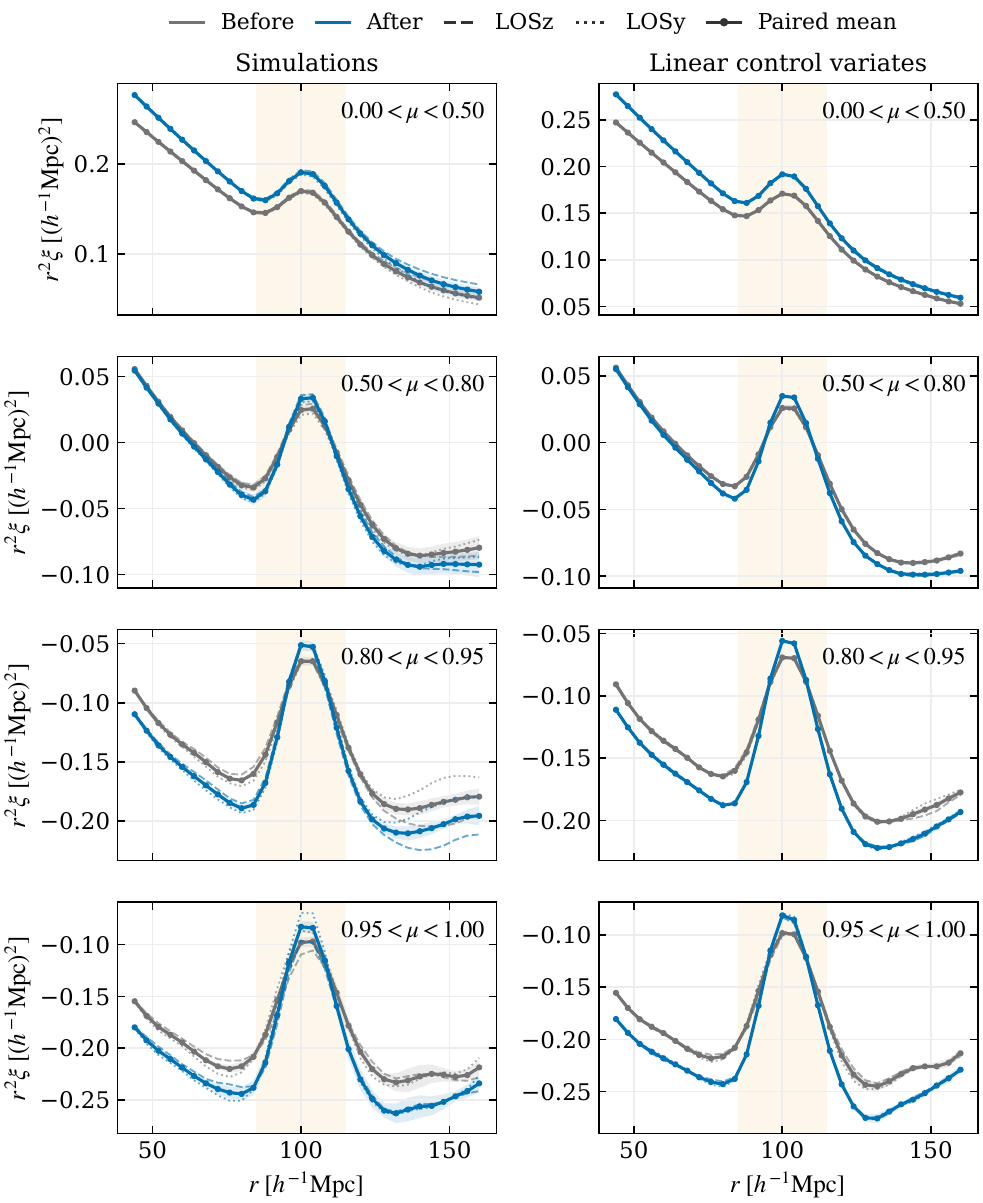}
  \caption{\textbf{Impact of BAO reconstruction in configuration space:} Same as Fig.~\ref{fig:wiggles} for the correlation function of Model~1 in four angular wedges, before (gray) and after (blue) reconstruction. The left column shows the raw simulations and the right column the linear control variates of Sec.~\ref{sec:lcv}. Dashed and dotted curves show the two lines of sight and solid lines their paired mean, with error bands from the scatter between the six phases. The shaded interval marks $85<r<115\Mpch$.}
  \label{fig:xi}
\end{figure*}

In Fig.~\ref{fig:wiggles} we show the measured BAO feature of Model~1 before (gray) and after (blue) reconstruction in two wedges. First, closest to the transverse direction (left, $0<\mu<0.2$) and, second, closest to the line-of-sight direction (right, $0.8<\mu<1$). Following baseline expectation, the BAO reconstruction mostly affects the radial scaling parameters manifesting in visibly sharpened oscillations from the second acoustic peak ($k\simeq0.14\hMpc$) onwards~\cite[see, Fig.~3 in][]{Chen2024BAOTheory}. This stems from redshift-space distortions that are enhanced along (or closer to) the line-of-sight displacements, $\Sigma_\parallel=(1+f)\,\Sigma_\perp$ in the Zel'dovich approximation~\cite{Chen2024BAOTheory, 2008PhRvD..78h3519M,Eisenstein2007}. In Fig.~\ref{fig:xi} we show the corresponding effect in configuration space for each angular wedge. We compare the ``raw'' simulation (left column) to the one where we apply the LCV technique (right column). 

We tabulate the key results for the fully sampled mock before and after reconstruction in Tab.~\ref{tab:precision}. For Model~1, the fitted radial (transverse) damping scale is reduced from $4.7$ (3.2) to $2.2\, (2.1)\Mpch$. The errors on the BAO scaling parameters $\apar$ and $\aperp$ shrink by $1.38$ and $1.20$ which correspond to gains in $\aiso$ and $\alphaap$ of $1.26$ and $1.32$, respectively. Whilst models 1 and 2 differ strongly in their redshift-space distortion parameter from models 3 and 4, the obtained gains from applying BAO reconstruction are somewhat similar for all four models. We find substantial gains for each BAO parameter. Since the two dilation parameters are anti-correlated with a correlation coefficient of $\simeq-0.4$, $\alphaap$ shows the biggest improvement. Note that we find gains that are smaller than for low-redshift galaxies as the bulk flows are smaller at $z=2.5$~\cite{Chen2024BAOTheory, deBelsunce:2024rvv}. After reconstruction, the fitted $\beta$ increases by about $0.1$ ($4$--$10\%$) in every model, because moving the samples does not exactly preserve the linear redshift-space anisotropy; since we leave $\beta_F$ as a free parameter the fit absorbs this change. Throughout this work, we use as smoothing radius $R=10\Mpch$ with the gains changing by less than $3\%$ between $R=8$ and $12.5\Mpch$.

\begin{table*}[t]
\begin{ruledtabular}
\begin{tabular}{ccccccccc}
Model & $(\Spar,\Sperp)$ pre $[\Mpch]$ & $(\Spar,\Sperp)$ post $[\Mpch]$
 & $100\,\sigma(\apar)$ & $100\,\sigma(\aperp)$
 & $G(\apar)$ & $G(\aperp)$ & $G(\aiso)$ & $G(\alphaap)$\\
\hline
1 & $4.7,\ 3.2$ & $2.2,\ 2.1$ & $0.29\to0.21$ & $0.16\to0.13$ & 1.38 & 1.20 & 1.26 & 1.32\\
2 & $4.7,\ 3.3$ & $2.2,\ 2.2$ & $0.29\to0.21$ & $0.16\to0.13$ & 1.37 & 1.20 & 1.26 & 1.32\\
3 & $4.9,\ 3.3$ & $2.8,\ 2.3$ & $0.29\to0.21$ & $0.15\to0.13$ & 1.34 & 1.17 & 1.22 & 1.29\\
4 & $5.0,\ 3.2$ & $2.8,\ 2.2$ & $0.29\to0.21$ & $0.15\to0.13$ & 1.35 & 1.17 & 1.23 & 1.29\\
\end{tabular}
\end{ruledtabular}
\caption{\textbf{Gains from BAO reconstruction:} Precision of the fully sampled forest before and after reconstruction (single fit to the mean of the six boxes). $\Spar$ and $\Sperp$ are the fitted damping scales, and $\sigma$ the profile-likelihood errors for the combined volume of the six boxes, $48\,h^{-3}{\rm Gpc}^3$, without noise. The gain is $G\equiv\sigma_{\rm pre}/\sigma_{\rm post}$. The dependence on the maximum wavenumber included in the fits is discussed in App.~\ref{app:BAO_kmax}.
}
\label{tab:precision}
\end{table*}

\begin{figure*}
  \includegraphics[width=\textwidth]{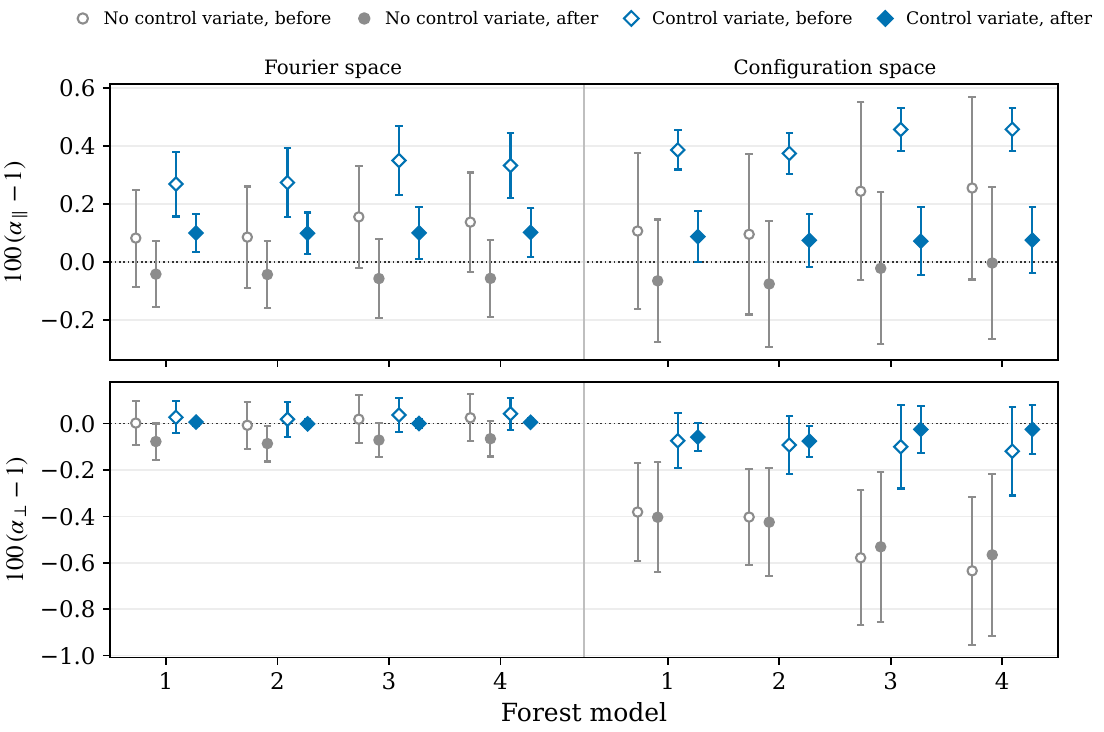}
  \caption{\textbf{BAO scaling parameters before and after reconstruction:} Offsets $100(\alpha-1)$ of the radial (top) and transverse (bottom panel) BAO scaling parameters before (open) and after (filled) reconstruction, for the four forest models in Fourier space (left) and in configuration space (right of the vertical line), without (gray) and with (blue) linear control variates. Each point averages the two lines of sight within each phase; the error bars are standard errors over the six phases. We use the baseline model of Sec.~\ref{sec:model} with the raw pre-reconstruction weights. 
  }
  \label{fig:bias}
\end{figure*}

An additional test of our methodology is to test for reconstruction-induced shifts in the BAO scale, using the baseline model and the paired responses of Sec.~\ref{sec:model}. In Fig.~\ref{fig:bias} we show the offsets of the scaling parameters from unity before (open) and after (filled) reconstruction. Without control variates, all offsets are consistent with zero. Applying LCVs, we find no transverse shift in per cent: $100(\aperp-1)=+0.02$ to $+0.04$ ($\pm0.07$) before and $-0.002$ to $+0.006$ ($\pm0.015$--$0.019$) after reconstruction in all models. In the radial direction, we measure $100(\apar-1)=+0.27$ to $+0.35$ ($\pm0.11$--$0.12$) before reconstruction, consistent with the non-linear shift of $+0.35\%$ measured in the same simulations~\cite{Hadzhiyska:2025cvk}. After reconstruction, we find a reduced shift of $100(\apar-1)=+0.10$, with errors of $0.07$--$0.09$ which are at the $1.1$--$1.5\sigma$-level consistent with no apparent shift. The paired responses are $100\Delta\apar=-0.17\pm0.08$, $-0.17\pm0.08$, $-0.25\pm0.07$, and $-0.23\pm0.06$ for Models~1--4 with reconstruction removing approximately two-thirds of the radial shift. We find consistent results for the configuration-space fits, introduced in Sec.~\ref{sec:model}. For Model 1 this yields $100(\apar-1)=+0.39\pm0.07\to+0.09\pm0.09$ and $100(\aperp-1)=-0.07\pm0.12\to-0.06\pm0.06$, with paired responses of $-0.30\pm0.09$ and $+0.02\pm0.09$. The large shift in radial direction stems from two realizations that yield a value of  $-0.8$ and $-1.2\%$ for each model. Applying LCVs we recover $-0.07\pm0.12\to-0.06\pm0.06$, which is consistent with no shift and the findings in Ref.~\cite{Hadzhiyska:2025cvk}.

\subsection{Applicability to observational data}
\label{sec:res_sparse}
\begin{figure}
    \includegraphics[width=\columnwidth]{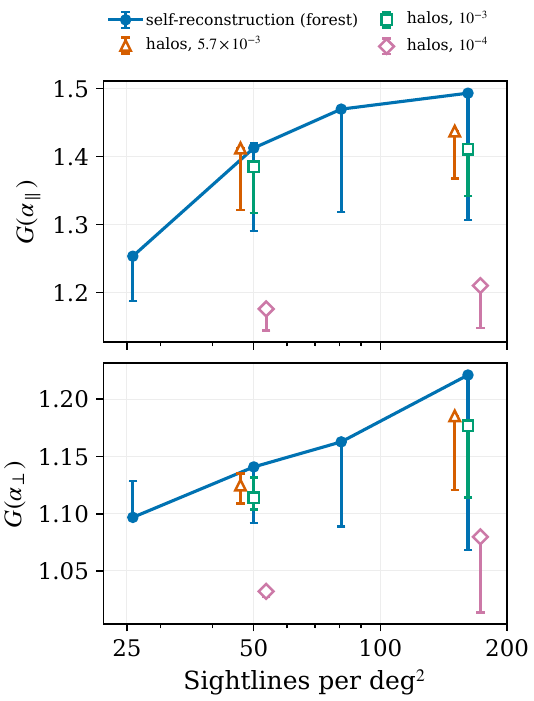}
    \caption{\textbf{Precision gains for varying sightline densities:} Precision gain, $G=\sigma_{\rm pre}/\sigma_{\rm post}$, for $\apar$ (top) and $\aperp$ (bottom) as a function of the sightline density, for self-reconstruction from the forest (Regime~III, line) and for displacements from external halo samples with three number densities (Regime~II, open symbols, slightly offset horizontally). Points are the central fits. The bars are not statistical errors: they span the three fit variants of Sec.~\ref{sec:sparse_est}, without the damping priors, with $k \leq 0.30 \hMpc$, and with the errors of each state rescaled by $\sqrt{\chi^2/{\rm dof}}$. The fully sampled forest, fitted in the same way, gives $1.38$ and $1.19$, matching Tab.~\ref{tab:precision}.
    }
    \label{fig:density}
  \end{figure}

To investigate the impact of a sparse window function on the BAO reconstruction, we apply two masks with a DESI-like and one with three times higher number density and compare the results using two methods to solve for the displacements. First, we use the \Lya forest itself (Regime~III) and, second, we use an external dark matter halo catalog with DESI-II-like number densities (Regime~II). In Fig.~\ref{fig:density} and Tab.~\ref{tab:sparse} we summarize our results. 

For a DESI-like density of $50\sqdeg$, regime III improves $\apar$ and $\aperp$ by factors of $1.41$ and $1.14$. At this density, the mean sightline separation of $9.9\Mpch$ is comparable to the smoothing radius, so that the smoothing kernel interpolates between neighboring sightlines (Sec.~\ref{sec:recon}). The obtained gain in radial direction matches the results from the fully sampled box. Increasing the number density of $160\sqdeg$ yields a gain of 1.49 with the transverse gain increasing by $1.22$. At $26\sqdeg$, where the mean sightline separation of $13.8\Mpch$ exceeds the smoothing radius, the radial (transverse) gain falls to $1.25$ ($1.10$). We caution the reader that the (surprisingly) good results from the masked case are driven by the fact that the line-of-sight modes are still intact, which assumes perfect knowledge of the continuum and no radial mask. Most of the information is lost in transverse direction instead of the radial direction. 

Using an external tracer in lieu of the \Lya skewers themselves yields comparable gains for large enough number densities. At $50\sqdeg$ with dark matter halos at a density of $\bar n=5.7\times10^{-3}$ and $10^{-3}\,h^3{\rm Mpc}^{-3}$ we improve $\apar$ by $1.41$ and $1.38$ and $\aperp$ by $1.13$ and $1.11$, respectively. Halos with $\bar n=10^{-4}\,h^3{\rm Mpc}^{-3}$ recover only $1.18$ and $1.03$, since their shot noise dominates the displacement on the scales that carry the BAO damping.
At $160\sqdeg$, the halo samples with $\bar n\geq10^{-3}\,h^3{\rm Mpc}^{-3}$ give $1.41$--$1.44$ and $1.18$--$1.19$. 

\begin{table*}[t]
  \begin{ruledtabular}
  \begin{tabular}{lcccc|cccc}
   & \multicolumn{4}{c}{$50\sqdeg$ (DESI-like)} & \multicolumn{4}{c}{$160\sqdeg$}\\
  Displacement source & $G(\apar)$ & $G(\aperp)$ & $G(\aiso)$ & $G(\alphaap)$
   & $G(\apar)$ & $G(\aperp)$ & $G(\aiso)$ & $G(\alphaap)$\\
  \hline
  Sightlines (Regime III) & 1.41 (1.29--1.42) & 1.14 (1.09--1.14) & 1.20 & 1.27 & 1.49 (1.31--1.49) & 1.22 (1.07--1.22) &
  1.30 & 1.40\\ 
  \textbf{Halos}&&&&&&&\\
  $\bar n=5.7\times10^{-3}\,h^3{\rm Mpc}^{-3}$ & 1.41 (1.32--1.41) & 1.13 (1.11--1.14) & 1.18 & 1.28 & 1.44
  (1.37--1.44) & 1.19 (1.12--1.19) & 1.27 & 1.34\\
  $\bar n=10^{-3}\,h^3{\rm Mpc}^{-3}$ & 1.38 (1.32--1.39) & 1.11 (1.10--1.13) & 1.17 & 1.25 & 1.41 (1.34--1.41) &
  1.18 (1.11--1.18) & 1.26 & 1.32\\
  $\bar n=10^{-4}\,h^3{\rm Mpc}^{-3}$ & 1.18 (1.14--1.18) & 1.03 (1.03--1.03) & 1.06 & 1.10 & 1.21 (1.15--1.21) &
  1.08 (1.01--1.08) & 1.12 & 1.17\\
  \end{tabular}
  \end{ruledtabular}
  \caption{\textbf{Precision gains for sparse sightlines:} Gains of Fig.~\ref{fig:density} at $50$ and $160\sqdeg$ for
  Model~1 (six-phase stack). The first number is the baseline fit; in parentheses the range over the three fit variants of
  Fig.~\ref{fig:density} (no damping prior, $k\leq 0.30\hMpc$, errors rescaled by $\sqrt{\chi^2/{\rm dof}}$), which mostly
  lower the gains. 
  $G(\aiso)$ and 
  $G(\alphaap)$ 
  are given for 
  the baseline fit.}
  \label{tab:sparse} 
  \end{table*}

The reduced $\chi^2$ of the fits is $1.1$--$1.2$ before and $1.2$--$1.3$ after reconstruction, and $1.45$ for the self-reconstruction at $160\sqdeg$. 
  We attribute this increase to the displaced pixels no longer lying on straight sightlines, which our currently used one-parameter aliasing template does not fully describe. The reconstruction displaces pixels in transverse direction, effectively smearing aliasing power in $k_\perp$-direction. This effect is most pronounced when doing sparse self-reconstruction at $160\sqdeg$, and absent in the fully sampled box, where no alias   term is needed and $\chi^2/{\rm dof}$ falls to $0.95$. 

  \begin{table}[t]
  \begin{ruledtabular}
  \begin{tabular}{lcccc}
   & \multicolumn{2}{c}{DESI ($50\sqdeg$)} & \multicolumn{2}{c}{DESI-II ($160\sqdeg$)}\\
  Fit & $G(\apar)$ & $G(\aperp)$ & $G(\apar)$ & $G(\aperp)$\\
  \hline
  \Lya auto & 1.42 & 1.13 & 1.49 & 1.22\\ 
  \Lya $\times$ QSO & 1.37 & 1.11 & 1.38 & 1.16\\ 
  Joint & 1.43 & 1.12 & 1.46 & 1.19\\
  \end{tabular} 
  \end{ruledtabular} 
  \caption{\textbf{Precision gain for the \Lya forest--quasar cross-correlation:} Model~1 with the forest on the $50$ and
  $160\sqdeg$ sightline masks and quasars (\texttt{CompaSO} halos) thinned to the DESI DR1 quasar density at $z=2.33$, $\bar
  n_Q=1.85\times10^{-5}\,h^3{\rm Mpc}^{-3}$, and to three times that for DESI-II. We compute the displacements using the \Lya forest for both   tracers with the same setup as detailed in Tab.~\ref{tab:sparse} without the damping prior. At the DESI density the errors on the cross-correlation before
  reconstruction ($0.97\%$ and $0.59\%$ for $\apar$ and $\aperp$) are twice those of the auto-correlation, so the joint fit does not improve over the auto-correlation.}
  \label{tab:cross}
  \end{table}

\subsection{BAO reconstruction for \Lya forest -- QSO cross-correlation} \label{sec:BAO_cross}
\begin{figure*}
    \centering
    \includegraphics[width=1\linewidth]{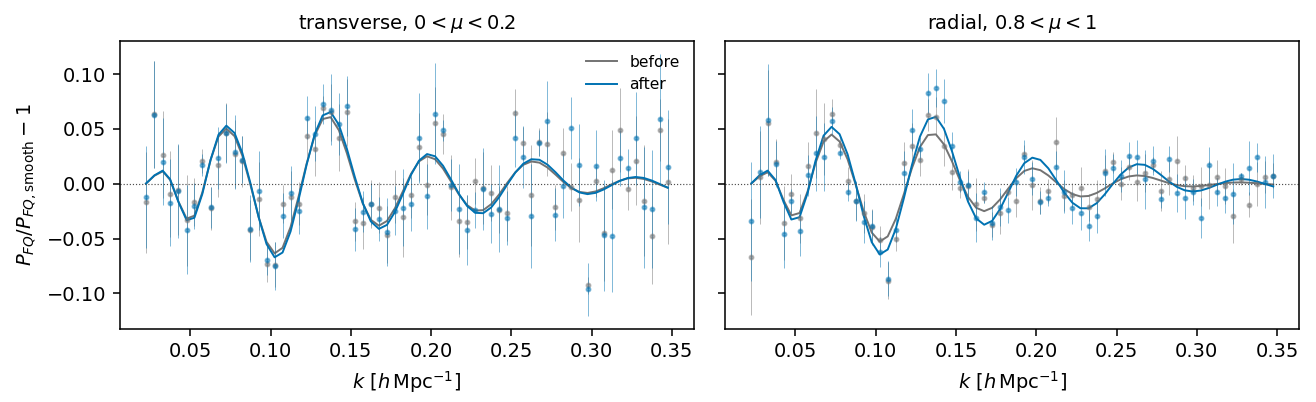}
    \caption{\textbf{BAO reconstruction for the cross-correlation:} Same as Fig.~\ref{fig:wiggles} but for the cross-correlation of the \Lya forest and a quasar catalog. Following baseline expectation, we find a marginal improvement in the transverse wedge with a noticeable improvement in the radial wedge. }
    \label{fig:wiggles_cross}
\end{figure*}

The cross-correlation of the \Lya forest with high-redshift quasars is a key ingredient to the high-redshift \Lya forest BAO constraints obtained by DESI~\cite{Font:2013, DESI2025DR2Lya}. Therefore, we apply the standard BAO reconstruction to the quasars (\texttt{CompaSO} halos) which are thinned to the DESI DR1 quasar density at $z=2.33$, $\bar n_Q=1.85\times10^{-5}\,h^3{\rm Mpc}^{-3}$, and to three times that for DESI-II. In Fig.~\ref{fig:wiggles_cross} we show the same plot as Fig.~\ref{fig:wiggles} but for the cross-correlation of the \Lya forest with quasars. We find a similar picture as for the auto-correlation, i.e.~most of the gain is in radial direction, and obtain for the cross-correlation an improvement of 1.37 on $\apar$ and 1.11 on $\aperp$. The results are summarized in Table~\ref{tab:cross}.

\section{Summary \& Conclusions}
\label{sec:conclusions}

Baryon acoustic oscillations (BAO) provide a robust probe of the expansion history of the Universe. Above redshift two the \Lya forest BAO measurements extend this probe into the matter-dominated era and complement the lower-redshift information from galaxy and quasar clustering~\cite{DESI2016,DESI2025DR2Lya}. The large influx of \Lya forest spectra observed by DESI and the forecasted cumulative precision of 0.2\% requires exquisite control over systematics affecting the BAO. These stem from the non-linearities in clustering damping and shifting the BAO feature. Recent progress in theoretical modeling within the one-loop EFT framework has demonstrated that it can unbias BAO measurements at the cost of a large number of EFT bias (or nuisance) parameters. To counter this, we propose to apply a non-linear transformation of the observed flux decrement to partially undo the non-linearities from clustering affecting the BAO signal -- a standard procedure for low-redshift galaxies~\cite[see, e.g.,~][for a review]{Chen2024BAOTheory}. This allows to improve cosmological distance measurements in the barely explored redshift regime $2\simlt z \simlt 4$. 

We have applied standard Zel'dovich BAO reconstruction to the \Lya forest and tested it on six large-volume \Lya forest mocks painted on top of the $N$-body simulation \textsc{AbacusSummit}. These result in four different \Lya forest models with varying magnitudes of the linear velocity-gradient bias parameter. We solve for the displacements using the ratio of the flux decrement to the linear bias, $\dF/b_F$, and move the flux pixels with the cosmological growth rate. Our main findings are:
\begin{enumerate}
    \item In the highly idealized scenario of a fully sampled \Lya forest mock, BAO reconstruction halves the radial  damping and tightens $\apar$, $\aperp$, $\aiso$, and $\alphaap$ by factors of  $1.34$--$1.40$, $1.17$--$1.22$, $1.22$--$1.30$, and $1.29$--$1.34$, respectively, across all four models spanning $\beta_F=1.0$--$2.4$.
    \item Using linear control variates to reduce the error bars by a factor of approximately $\sqrt{50}$, we find that BAO reconstruction reduces the shift in the radial direction from $+0.3\%$ to $+0.1\%$ and leaves the transverse parameter unbiased. 
    \item We perform BAO reconstruction that is directly applicable to realistic surveys and test a DESI-like number density of $50\sqdeg$ resulting in  improvements on $\apar$, $\aperp$, $\aiso$, and $\alphaap$ by $1.41$, $1.14$, $1.20$,  and $1.27$. When using an external dark matter halo tracer to compute the displacements with a futuristic number density of $\bar n\ge10^{-3}\,h^3{\rm Mpc}^{-3}$ we reach gains up to $1.4$--$1.5$ and $1.2$ in radial and transverse direction, respectively. Decreasing the number density to $\bar{n}=10^{-4}\,h^3{\rm Mpc}^{-3}$ improves the BAO scaling parameters at the 3--10\% level. 
    \item We apply the BAO reconstruction to the auto-correlation of the \Lya forest and to the cross-correlation of the \Lya forest with quasars and find for both similar improvements, even in the presence of a DESI-like survey mask. 
\end{enumerate}
Reconstruction is a promising way to tighten cosmological distance measurements from the BAO feature with DESI and its successor DESI-II. We emphasize, however, that \textit{we show its promise in a highly idealized scenario}. Whilst we test realistic sightline densities, we ignore a plethora of data-specific effects: the instrument response and pixel noise, continuum fitting, finite forest lengths and gaps in the spectra, sky residuals, an anisotropic mask and selection function, the redshift evolution of the spectra, and high column density systems and metals, both of which can leak power into high-$k$ bins near the BAO scale. Pixel noise and continuum-fitting distortions~\cite{Busca2025Continuum} degrade both the high-$k$ BAO information that reconstruction restores and, in Regime~III, the displacement estimate, although external tracers are helpful for averting these issues. Continuum fitting will likely have the strongest impact, as current methods~\cite{DESI_lya_2024} project out long-wavelength modes along each sightline which are crucial for BAO reconstruction. Our gains are thus to be understood as upper limits. We have demonstrated a mitigation strategy for the displacement estimate: an external tracer with $\bar n\geq10^{-3}\,h^3{\rm Mpc}^{-3}$, whose displacement does not depend on the observed flux, gives gains comparable to self-reconstruction (Sec.~\ref{sec:res_sparse}). Our tracers are, however, randomly thinned halos with a known bias, without a realistic galaxy selection function or redshift errors. Further, our covariances are approximate: we use a rescaled eBOSS DR16 covariance for the correlation function~\cite{dMdB:2020, Hadzhiyska:2025cvk} and Gaussian analytic covariances for the power spectrum. These observational effects can be tested with readily available \texttt{LyaCoLoRe} simulations~\cite{Farr:2019xij, Ramirez-Perez:2021cpq, Bernal:2026klr,DESI:2025qqu}.\footnote{In particular, the \texttt{CoLoRe-2LPT} \Lya forest mocks yield a consistent value of the BAO shift to the present mocks of 0.3\% in radial direction~\cite{Bernal:2026klr}.} Finally, our six realizations share a single cosmology, so we do not test how reconstruction depends on the cosmological parameters. Another fruitful avenue is to add the cross-correlation with high-redshift quasars, which carries a large part of the constraining power of current \Lya forest BAO analyses~\cite{Font:2013, DESI_lya_2024, Chudaykin:2025gsh}. We leave these avenues to future work.

\begin{acknowledgments}
We thank Daniel Eisenstein, Julien Guy, and Andreu Font-Ribera for useful discussions especially during the early stages of this project. We are especially grateful to the \Lya forest enthusiasts Misha Ivanov and Jamie Sullivan for fruitful discussions and Eric Armengaud for comments on the draft. 

This work is supported by the National Science Foundation under Cooperative Agreement PHY-2019786 (The NSF AI Institute for Artificial Intelligence and Fundamental Interactions, \url{http://iaifi.org/}). 
This research used resources of the National Energy Research Scientific Computing Center (NERSC), a U.S. Department of Energy Office of Science User Facility operated under Contract No. DE-AC02-05CH11231.

\textit{AI usage statement:} This work was entirely conceptualized and written by the authors. Claude (Opus 5.5 and Fable 5.1; Anthropic) was used, in parts, to perform numerical computations and for code optimization in \texttt{python} as well as to edit the manuscript for clarity. The authors have checked and validated all the results, taking full responsibility for all results presented herein.
\end{acknowledgments}

\section*{Data availability}
We make all our synthetic maps and catalogs publicly available on Globus through NERSC SHARE at this link: \url{https://app.globus.org/file-manager?origin_id=9ce29982-eed1-11ed-9bb4-c9bb788c490e&path=\%2F} under the name ``AbacusSummit Lyman Alpha Forest''. 

\appendix

\section{BAO reconstruction as a function of wavenumber} \label{app:BAO_kmax}
\begin{figure*}
  \includegraphics[width=\textwidth]{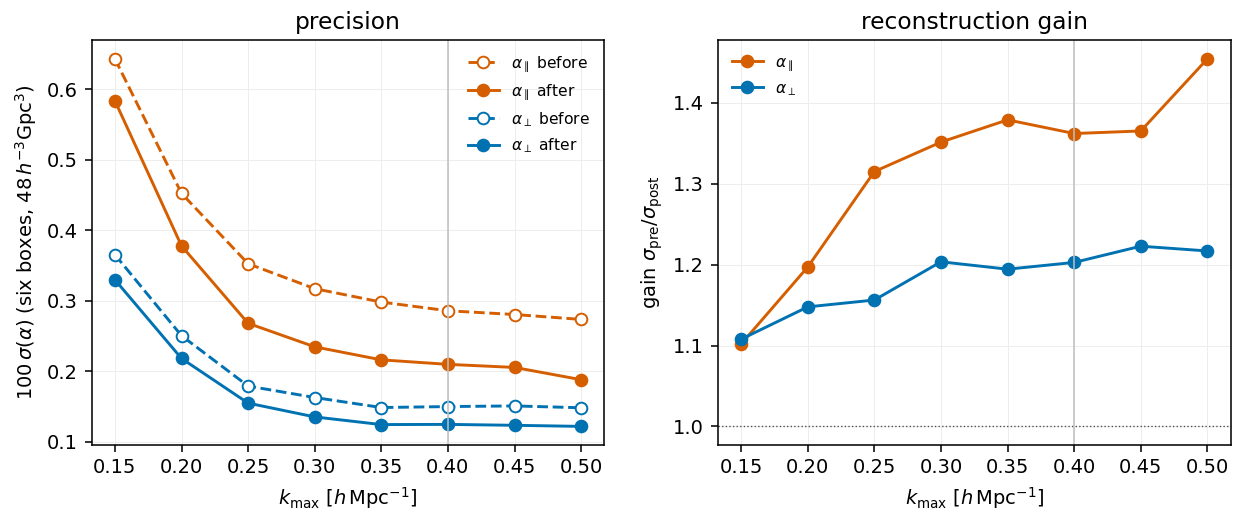}
  \caption{\textbf{BAO reconstruction as a function of wavenumber:} We show the evolution of the BAO precision (left panel) and the gain (right panel). We find consistent improvement in the BAO scaling parameters with increasing wavenumber up to $k_{\rm max}\simeq 0.3\hMpc$, using smaller scales in the fits only marginally adds more information. For the gain we find a steeper gain for the radial than for the transverse scaling parameters. 
  }
  \label{fig:BAO_kmax}
\end{figure*}

In Fig.~\ref{fig:BAO_kmax} we compare the precision and gain from the BAO reconstruction as a function of wavenumber for model 1. We find a consistent behavior for both BAO scaling parameters: up to $k_{\rm max}\simeq 0.3\hMpc$ we benefit from a consistent increase and when including smaller scales we only find marginal improvements. 

\section{Choice of smoothing radius}
\label{app:smoothing}

\begin{figure}
  \includegraphics[width=0.9\columnwidth]{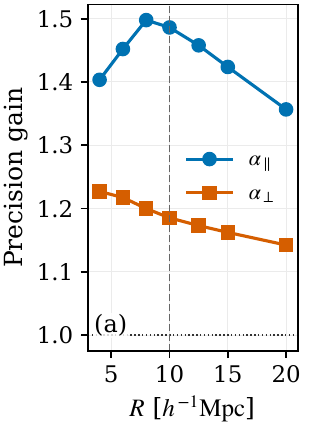}
  \caption{\textbf{Impact of smoothing radius on gain:} Variation of the smoothing radius in the range $4 \leq R\, [\hMpc] \leq 20$ for Model~1, phase 000 and line-of-sight LOSz. The dotted horizontal line denotes the reference value pre-reconstruction and the dashed vertical line indicates our smoothing radius choice.}
  \label{fig:smoothing}
\end{figure}

We investigate the impact of varying the smoothing scale in Fig.~\ref{fig:smoothing} in the range from $4$ to 
$20\Mpch$ for Model~1 for one realization (phase one) in one phase. The radial gain has a broad maximum at $R=8$--$12.5\Mpch$ and falls to $1.36$ at $R=20\Mpch$ while the transverse gain
decreases slowly with smoothing radius, from $1.23$ to $1.14$. We adopt
$R=10\Mpch$ which maximizes the precision gain in radial direction with an intermediate level of loss in transverse direction. 

\bibliographystyle{aux_files/JHEP}
\bibliography{references}

@article{Weinberg2013,
  author = {Weinberg, David H. and Mortonson, Michael J. and Eisenstein, Daniel J. and Hirata, Christopher and Riess, Adam G. and Rozo, Eduardo},
  title = {Observational Probes of Cosmic Acceleration},
  journal = {Physics Reports},
  volume = {530},
  pages = {87--255},
  year = {2013},
  doi = {10.1016/j.physrep.2013.05.001},
  eprint = {1201.2434},
  archivePrefix = {arXiv}
}

@article{Eisenstein2005,
  author = {Eisenstein, Daniel J. and others},
  title = {Detection of the Baryon Acoustic Peak in the Large-Scale Correlation Function of {SDSS} Luminous Red Galaxies},
  journal = {Astrophysical Journal},
  volume = {633},
  pages = {560--574},
  year = {2005},
  doi = {10.1086/466512},
  eprint = {astro-ph/0501171},
  archivePrefix = {arXiv}
}

@article{Cole2005,
  author = {Cole, Shaun and others},
  title = {The 2dF Galaxy Redshift Survey: Power-Spectrum Analysis of the Final Data Set and Cosmological Implications},
  journal = {Monthly Notices of the Royal Astronomical Society},
  volume = {362},
  pages = {505--534},
  year = {2005},
  doi = {10.1111/j.1365-2966.2005.09318.x},
  eprint = {astro-ph/0501174},
  archivePrefix = {arXiv}
}

@article{EisensteinSeoWhite2007,
  author = {Eisenstein, Daniel J. and Seo, Hee-Jong and White, Martin J.},
  title = {On the Robustness of the Acoustic Scale in the Low-Redshift Clustering of Matter},
  journal = {Astrophysical Journal},
  volume = {664},
  pages = {660--674},
  year = {2007},
  doi = {10.1086/518755},
  eprint = {astro-ph/0604361},
  archivePrefix = {arXiv}
}

@article{Eisenstein2007,
  author = {Eisenstein, Daniel J. and Seo, Hee-Jong and Sirko, Edwin and Spergel, David N.},
  title = {Improving Cosmological Distance Measurements by Reconstruction of the Baryon Acoustic Peak},
  journal = {Astrophysical Journal},
  volume = {664},
  pages = {675--679},
  year = {2007},
  doi = {10.1086/518712},
  eprint = {astro-ph/0604362},
  archivePrefix = {arXiv}
}

@article{CrocceScoccimarro2008,
  author = {Crocce, Martin and Scoccimarro, Roman},
  title = {Nonlinear Evolution of Baryon Acoustic Oscillations},
  journal = {Physical Review D},
  volume = {77},
  pages = {023533},
  year = {2008},
  doi = {10.1103/PhysRevD.77.023533},
  eprint = {0704.2783},
  archivePrefix = {arXiv}
}

@article{PadmanabhanWhite2009,
  author = {Padmanabhan, Nikhil and White, Martin},
  title = {Calibrating the Baryon Oscillation Ruler for Matter and Halos},
  journal = {Physical Review D},
  volume = {80},
  pages = {063508},
  year = {2009},
  doi = {10.1103/PhysRevD.80.063508},
  eprint = {0906.1198},
  archivePrefix = {arXiv}
}

@article{SherwinZaldarriaga2012,
  author = {Sherwin, Blake D. and Zaldarriaga, Matias},
  title = {Shift of the Baryon Acoustic Oscillation Scale: A Simple Physical Picture},
  journal = {Physical Review D},
  volume = {85},
  pages = {103523},
  year = {2012},
  doi = {10.1103/PhysRevD.85.103523},
  eprint = {1202.3998},
  archivePrefix = {arXiv}
}

@article{ChenVlahWhite2019,
  author = {Chen, Shi-Fan and Vlah, Zvonimir and White, Martin},
  title = {The Reconstructed Power Spectrum in the {Zeldovich} Approximation},
  journal = {Journal of Cosmology and Astroparticle Physics},
  volume = {2019},
  number = {09},
  pages = {017},
  year = {2019},
  doi = {10.1088/1475-7516/2019/09/017},
  eprint = {1907.00043},
  archivePrefix = {arXiv}
}

@article{Chen2024Recon,
  author = {Chen, X. and others},
  title = {Extensive Analysis of Reconstruction Algorithms for {DESI} 2024 Baryon Acoustic Oscillations},
  journal = {Journal of Cosmology and Astroparticle Physics},
  volume = {2026},
  number = {05},
  pages = {001},
  year = {2026},
  doi = {10.1088/1475-7516/2026/05/001},
  eprint = {2411.19738},
  archivePrefix = {arXiv}
}

@article{Chen2024BAOTheory,
  author = {Chen, Shi-Fan and Howlett, Cullan and White, Martin and McDonald, Patrick and Ross, Ashley J. and Seo, Hee-Jong and Padmanabhan, Nikhil and others},
  title = {Baryon Acoustic Oscillation Theory and Modelling Systematics for the {DESI} 2024 results},
  journal = {Monthly Notices of the Royal Astronomical Society},
  volume = {534},
  number = {1},
  pages = {544--574},
  year = {2024},
  doi = {10.1093/mnras/stae2090},
  eprint = {2402.14070},
  archivePrefix = {arXiv}
}

@article{DESI2016,
  author = {{DESI Collaboration}},
  title = {The {DESI} Experiment Part {I}: Science, Targeting, and Survey Design},
  journal = {arXiv e-prints},
  year = {2016},
  eprint = {1611.00036},
  archivePrefix = {arXiv}
}

@article{SavitzkyGolay1964,
  author = {Savitzky, Abraham and Golay, Marcel J. E.},
  title = {Smoothing and Differentiation of Data by Simplified Least Squares Procedures},
  journal = {Analytical Chemistry},
  volume = {36},
  pages = {1627--1639},
  year = {1964},
  doi = {10.1021/ac60214a047}
}

@article{Maksimova2021,
  author = {Maksimova, Nina A. and Garrison, Lehman H. and Eisenstein, Daniel J. and Hadzhiyska, Boryana and Bose, Sownak and Satterthwaite, Thomas P.},
  title = {{AbacusSummit}: A Massive Set of High-Accuracy, High-Resolution $N$-Body Simulations},
  journal = {Monthly Notices of the Royal Astronomical Society},
  volume = {508},
  pages = {4017--4037},
  year = {2021},
  doi = {10.1093/mnras/stab2484},
  eprint = {2110.11398},
  archivePrefix = {arXiv}
}

@article{Kokron2022,
  author = {Kokron, Nickolas and Chen, Shi-Fan and White, Martin and DeRose, Joseph and Maus, Mark},
  title = {Accurate Predictions from Small Boxes: Variance Suppression via the {Zel'dovich} Approximation},
  journal = {Journal of Cosmology and Astroparticle Physics},
  volume = {2022},
  number = {09},
  pages = {059},
  year = {2022},
  doi = {10.1088/1475-7516/2022/09/059},
  eprint = {2205.15327},
  archivePrefix = {arXiv}
}

@article{DeRose2023,
  author = {DeRose, Joseph and Chen, Shi-Fan and Kokron, Nickolas and White, Martin},
  title = {Precision Redshift-Space Galaxy Power Spectra Using Zel'dovich Control Variates},
  journal = {Journal of Cosmology and Astroparticle Physics},
  volume = {2023},
  number = {02},
  pages = {008},
  year = {2023},
  doi = {10.1088/1475-7516/2023/02/008},
  eprint = {2210.14239},
  archivePrefix = {arXiv}
}

@article{HadzhiyskaCV2023,
  author = {Hadzhiyska, Boryana and others},
  title = {Mitigating the Noise of {DESI} Mocks Using Analytic Control Variates},
  journal = {The Open Journal of Astrophysics},
  volume = {6},
  pages = {38},
  year = {2023},
  doi = {10.21105/astro.2308.12343},
  eprint = {2308.12343},
  archivePrefix = {arXiv}
}

@article{AlcockPaczynski1979,
  author = {Alcock, Charles and Paczynski, Bohdan},
  title = {An Evolution Free Test for Non-Zero Cosmological Constant},
  journal = {Nature},
  volume = {281},
  pages = {358--359},
  year = {1979},
  doi = {10.1038/281358a0}
}

@article{EisensteinHu1998,
  author = {Eisenstein, Daniel J. and Hu, Wayne},
  title = {Baryonic Features in the Matter Transfer Function},
  journal = {Astrophysical Journal},
  volume = {496},
  pages = {605--614},
  year = {1998},
  doi = {10.1086/305424},
  eprint = {astro-ph/9709112},
  archivePrefix = {arXiv}
}

@article{Padmanabhan2012,
  author = {Padmanabhan, Nikhil and others},
  title = {A 2 Per Cent Distance to $z=0.35$ by Reconstructing Baryon Acoustic Oscillations. {I}. Methods and Application to the {Sloan Digital Sky Survey}},
  journal = {Monthly Notices of the Royal Astronomical Society},
  volume = {427},
  pages = {2132--2145},
  year = {2012},
  doi = {10.1111/j.1365-2966.2012.21888.x},
  eprint = {1202.0090},
  archivePrefix = {arXiv}
}

@article{Seo2016,
  author = {Seo, Hee-Jong and Beutler, Florian and Ross, Ashley J. and Saito, Shun},
  title = {Modeling the Reconstructed {BAO} in Fourier Space},
  journal = {Monthly Notices of the Royal Astronomical Society},
  volume = {460},
  pages = {2453--2471},
  year = {2016},
  doi = {10.1093/mnras/stw1138},
  eprint = {1511.00663},
  archivePrefix = {arXiv}
}

@article{Burden2014,
  author = {Burden, Angela and Percival, Will J. and Manera, Marc and Cuesta, Antonio J. and Vargas Maga{\~n}a, Mariana and Ho, Shirley},
  title = {Efficient Reconstruction of Linear Baryon Acoustic Oscillations in Galaxy Surveys},
  journal = {Monthly Notices of the Royal Astronomical Society},
  volume = {445},
  pages = {3152--3168},
  year = {2014},
  doi = {10.1093/mnras/stu1965},
  eprint = {1408.1348},
  archivePrefix = {arXiv}
}

@article{Paillas2025,
  author = {Paillas, Enrique and others},
  title = {Optimal Reconstruction of Baryon Acoustic Oscillations for {DESI} 2024},
  journal = {Journal of Cosmology and Astroparticle Physics},
  volume = {2025},
  number = {01},
  pages = {142},
  year = {2025},
  doi = {10.1088/1475-7516/2025/01/142},
  eprint = {2404.03005},
  archivePrefix = {arXiv}
}

@article{Busca2025Continuum,
  author = {Busca, Nicolas and others},
  title = {The Effects of Continuum Fitting on Lyman-Alpha Forest Correlations},
  journal = {Journal of Cosmology and Astroparticle Physics},
  volume = {2025},
  number = {09},
  pages = {020},
  year = {2025},
  doi = {10.1088/1475-7516/2025/09/020},
  eprint = {2506.15262},
  archivePrefix = {arXiv}
}

@article{DESI2025DR2Cosmo,
  author = {{DESI Collaboration}},
  title = {{DESI} {DR2} Results. {II}. Measurements of Baryon Acoustic Oscillations and Cosmological Constraints},
  journal = {Physical Review D},
  volume = {112},
  number = {8},
  pages = {083515},
  year = {2025},
  doi = {10.1103/tr6y-kpc6},
  eprint = {2503.14738},
  archivePrefix = {arXiv}
}

@article{Schlegel2022Roadmap,
  author = {Schlegel, David J. and Ferraro, Simone and Aldering, Greg and Baltay, Charles and BenZvi, Segev and others},
  title = {A Spectroscopic Road Map for Cosmic Frontier: {DESI}, {DESI-II}, {Stage-5}},
  journal = {arXiv e-prints},
  year = {2022},
  doi = {10.2172/1887254},
  eprint = {2209.03585},
  archivePrefix = {arXiv},
  note = {Contribution to Snowmass 2021}
}

@article{SpecS5_2025,
  author = {Besuner, Robert and Dey, Arjun and Drlica-Wagner, Alex and Ebina, Haruki and Fernandez Moroni, Guillermo and others},
  collaboration = {Spec-S5},
  title = {The Spectroscopic Stage-5 Experiment},
  journal = {arXiv e-prints},
  year = {2025},
  eprint = {2503.07923},
  archivePrefix = {arXiv}
}

@article{Ravi2024LAE,
  author = {Ravi, Jyotsna and Hadzhiyska, Boryana and White, Martin J. and Hernquist, Lars and Bose, Sownak},
  title = {Examining {Lyman}-alpha Emitters through Simulations in Anticipation of the {DESI-II} Survey},
  journal = {Physical Review D},
  volume = {110},
  number = {10},
  pages = {103509},
  year = {2024},
  doi = {10.1103/PhysRevD.110.103509},
  eprint = {2403.02414},
  archivePrefix = {arXiv}
}

@article{Payerne2025LBG,
  author = {Payerne, Constantin and d'Assignies Doumerg, William and Y{\`e}che, Christophe and Ruhlmann-Kleider, Vanina and Raichoor, Anand and others},
  title = {Selection of High-Redshift {Lyman}-Break Galaxies from Broadband and Wide Photometric Surveys},
  journal = {Journal of Cosmology and Astroparticle Physics},
  volume = {2025},
  number = {05},
  pages = {031},
  year = {2025},
  doi = {10.1088/1475-7516/2025/05/031},
  eprint = {2410.08062},
  archivePrefix = {arXiv}
}

@article{Lee2014Requirements,
  author = {Lee, Khee-Gan and Hennawi, Joseph F. and White, Martin and Croft, Rupert and Ozbek, Melih},
  title = {Observational Requirements for {Ly$\alpha$} Forest Tomographic Mapping of Large-Scale Structure at $z\sim2$},
  journal = {Astrophysical Journal},
  volume = {788},
  number = {1},
  pages = {49},
  year = {2014},
  doi = {10.1088/0004-637X/788/1/49},
  eprint = {1309.1477},
  archivePrefix = {arXiv}
}

@article{Horowitz2019,
  author = {Horowitz, Benjamin and Lee, Khee-Gan and White, Martin and Krolewski, Alex and Ata, Metin},
  title = {{TARDIS}. {I}. A Constrained Reconstruction Approach to Modeling the $z\sim2.5$ Cosmic Web Probed by {Ly$\alpha$} Forest Tomography},
  journal = {Astrophysical Journal},
  volume = {887},
  number = {1},
  pages = {61},
  year = {2019},
  doi = {10.3847/1538-4357/ab4d4c},
  eprint = {1903.09049},
  archivePrefix = {arXiv}
}

@article{Ravoux2020,
  author = {Ravoux, C. and Armengaud, E. and Walther, M. and Etourneau, T. and Pomar{\`e}de, D. and others},
  title = {A Tomographic Map of the Large-Scale Matter Distribution Using the {eBOSS}--{Stripe} 82 {Ly$\alpha$} Forest},
  journal = {Journal of Cosmology and Astroparticle Physics},
  volume = {2020},
  number = {07},
  pages = {010},
  year = {2020},
  doi = {10.1088/1475-7516/2020/07/010},
  eprint = {2004.01448},
  archivePrefix = {arXiv}
}

@article{Horowitz2021,
  author = {Horowitz, Benjamin and Zhang, Benjamin and Lee, Khee-Gan and Kooistra, Robin},
  title = {{TARDIS}. {II}. Synergistic Density Reconstruction from {Ly$\alpha$} Forest and Spectroscopic Galaxy Surveys with Applications to Protoclusters and the Cosmic Web},
  journal = {Astrophysical Journal},
  volume = {906},
  number = {2},
  pages = {110},
  year = {2021},
  doi = {10.3847/1538-4357/abca35},
  eprint = {2007.15994},
  archivePrefix = {arXiv}
}

@article{deBelsunce2026FieldLevel,
    author = "de Belsunce, Roger and Ivanov, Mikhail M. and Sullivan, James M. and Akitsu, Kazuyuki and Chen, Shi-Fan",
    title = "{Modeling the Cosmological Lyman-{\ensuremath{\alpha}} Forest at the Field Level}",
    eprint = "2507.00284",
    archivePrefix = "arXiv",
    primaryClass = "astro-ph.CO",
    reportNumber = "MIT-CTP/5882",
    doi = "10.1103/8g9h-68ms",
    journal = "Phys. Rev. Lett.",
    volume = "136",
    number = "10",
    pages = "101001",
    year = "2026"
}

@misc{Pyrecon,
  author = {{pyrecon developers}},
  title = {{pyrecon}: package for {BAO} reconstruction},
  year = {2026},
  howpublished = {\url{https://github.com/cosmodesi/pyrecon}},
  note = {Accessed September 2026}
}

@article{deBelsunce:2024knf,
    author = "de Belsunce, Roger and Philcox, Oliver H. E. and Irsic, Vid and McDonald, Patrick and Guy, Julien and Palanque-Delabrouille, Nathalie",
    title = "{The 3D Lyman-\ensuremath{\alpha} forest power spectrum from eBOSS DR16}",
    eprint = "2403.08241",
    archivePrefix = "arXiv",
    primaryClass = "astro-ph.CO",
    doi = "10.1093/mnras/stae2035",
    journal = "Mon. Not. Roy. Astron. Soc.",
    volume = "533",
    number = "3",
    pages = "3756--3770",
    year = "2024"
}

@ARTICLE{DESI_lya_2024,
       author = {{DESI Collaboration} and {Adame}, A.~G. and {Aguilar}, J. and {Ahlen}, S. and {Alam}, S. and {Alexander}, D.~M. and {Alvarez}, M. and {Alves}, O. and {Anand}, A. and {Andrade}, U. and {Armengaud}, E. and {Avila}, S. and {Aviles}, A. and {Awan}, H. and {Bailey}, S. and {Baltay}, C. and {Bault}, A. and {Bautista}, J. and {Behera}, J. and {BenZvi}, S. and {Beutler}, F. and {Bianchi}, D. and {Blake}, C. and {Blum}, R. and {Brieden}, S. and {Brodzeller}, A. and {Brooks}, D. and {Buckley-Geer}, E. and {Burtin}, E. and {Calderon}, R. and {Canning}, R. and {Carnero Rosell}, A. and {Cereskaite}, R. and {Cervantes-Cota}, J.~L. and {Chabanier}, S. and {Chaussidon}, E. and {Chaves-Montero}, J. and {Chen}, S. and {Chen}, X. and {Claybaugh}, T. and {Cole}, S. and {Cuceu}, A. and {Davis}, T.~M. and {Dawson}, K. and {de la Cruz}, R. and {de la Macorra}, A. and {de Mattia}, A. and {Deiosso}, N. and {Dey}, A. and {Dey}, B. and {Ding}, J. and {Ding}, Z. and {Doel}, P. and {Edelstein}, J. and {Eftekharzadeh}, S. and {Eisenstein}, D.~J. and {Elliott}, A. and {Fagrelius}, P. and {Fanning}, K. and {Ferraro}, S. and {Ereza}, J. and {Findlay}, N. and {Flaugher}, B. and {Font-Ribera}, A. and {Forero-S{\'a}nchez}, D. and {Forero-Romero}, J.~E. and {Garcia-Quintero}, C. and {Gazta{\~n}aga}, E. and {Gil-Mar{\'\i}n}, H. and {Gontcho}, S. Gontcho A and {Gonzalez-Morales}, A.~X. and {Gonzalez-Perez}, V. and {Gordon}, C. and {Green}, D. and {Gruen}, D. and {Gsponer}, R. and {Gutierrez}, G. and {Guy}, J. and {Hadzhiyska}, B. and {Hahn}, C. and {Hanif}, M.~M. S and {Herrera-Alcantar}, H.~K. and {Honscheid}, K. and {Howlett}, C. and {Huterer}, D. and {Ir{\v{s}}i{\v{c}}}, V. and {Ishak}, M. and {Juneau}, S. and {Kara{\c{c}}ayli}, N.~G. and {Kehoe}, R. and {Kent}, S. and {Kirkby}, D. and {Kremin}, A. and {Krolewski}, A. and {Lai}, Y. and {Lan}, T. -W. and {Landriau}, M. and {Lang}, D. and {Lasker}, J. and {Le Goff}, J.~M. and {Le Guillou}, L. and {Leauthaud}, A. and {Levi}, M.~E. and {Li}, T.~S. and {Linder}, E. and {Lodha}, K. and {Magneville}, C. and {Manera}, M. and {Margala}, D. and {Martini}, P. and {Maus}, M. and {McDonald}, P. and {Medina-Varela}, L. and {Meisner}, A. and {Mena-Fern{\'a}ndez}, J. and {Miquel}, R. and {Moon}, J. and {Moore}, S. and {Moustakas}, J. and {Mudur}, N. and {Mueller}, E. and {Mu{\~n}oz-Guti{\'e}rrez}, A. and {Myers}, A.~D. and {Nadathur}, S. and {Napolitano}, L. and {Neveux}, R. and {Newman}, J.~A. and {Nguyen}, N.~M. and {Nie}, J. and {Niz}, G. and {Noriega}, H.~E. and {Padmanabhan}, N. and {Paillas}, E. and {Palanque-Delabrouille}, N. and {Pan}, J. and {Penmetsa}, S. and {Percival}, W.~J. and {Pieri}, M.~M. and {Pinon}, M. and {Poppett}, C. and {Porredon}, A. and {Prada}, F. and {P{\'e}rez-Fern{\'a}ndez}, A. and {P{\'e}rez-R{\`a}fols}, I. and {Rabinowitz}, D. and {Raichoor}, A. and {Ram{\'\i}rez-P{\'e}rez}, C. and {Ramirez-Solano}, S. and {Rashkovetskyi}, M. and {Ravoux}, C. and {Rezaie}, M. and {Rich}, J. and {Rocher}, A. and {Rockosi}, C. and {Roe}, N.~A. and {Rosado-Marin}, A. and {Ross}, A.~J. and {Rossi}, G. and {Ruggeri}, R. and {Ruhlmann-Kleider}, V. and {Samushia}, L. and {Sanchez}, E. and {Saulder}, C. and {Schlafly}, E.~F. and {Schlegel}, D. and {Schubnell}, M. and {Seo}, H. and {Sharples}, R. and {Silber}, J. and {Sinigaglia}, F. and {Slosar}, A. and {Smith}, A. and {Sprayberry}, D. and {Tan}, T. and {Tarl{\'e}}, G. and {Trusov}, S. and {Vaisakh}, R. and {Valcin}, D. and {Valdes}, F. and {Vargas-Maga{\~n}a}, M. and {Verde}, L. and {Walther}, M. and {Wang}, B. and {Wang}, M.~S. and {Weaver}, B.~A. and {Weaverdyck}, N. and {Wechsler}, R.~H. and {Weinberg}, D.~H. and {White}, M. and {Yu}, J. and {Yu}, Y. and {Yuan}, S. and {Y{\`e}che}, C. and {Zaborowski}, E.~A. and {Zarrouk}, P. and {Zhang}, H. and {Zhao}, C. and {Zhao}, R. and {Zhou}, R. and {Zou}, H.},
        title = "{DESI 2024 IV: Baryon Acoustic Oscillations from the Lyman Alpha Forest}",
      journal = {arXiv e-prints},
         year = 2024,
        month = apr,
          eid = {arXiv:2404.03001},
        pages = {arXiv:2404.03001},
          doi = {10.48550/arXiv.2404.03001},
archivePrefix = {arXiv},
       eprint = {2404.03001},
 primaryClass = {astro-ph.CO},
       adsurl = {https://ui.adsabs.harvard.edu/abs/2024arXiv240403001D}
}

@ARTICLE{DESI_BAO_2024,
       author = {{DESI Collaboration} and {Adame}, A.~G. and {Aguilar}, J. and {Ahlen}, S. and {Alam}, S. and {Alexander}, D.~M. and {Alvarez}, M. and {Alves}, O. and {Anand}, A. and {Andrade}, U. and {Armengaud}, E. and {Avila}, S. and {Aviles}, A. and {Awan}, H. and {Bahr-Kalus}, B. and {Bailey}, S. and {Baltay}, C. and {Bault}, A. and {Behera}, J. and {BenZvi}, S. and {Bera}, A. and {Beutler}, F. and {Bianchi}, D. and {Blake}, C. and {Blum}, R. and {Brieden}, S. and {Brodzeller}, A. and {Brooks}, D. and {Buckley-Geer}, E. and {Burtin}, E. and {Calderon}, R. and {Canning}, R. and {Carnero Rosell}, A. and {Cereskaite}, R. and {Cervantes-Cota}, J.~L. and {Chabanier}, S. and {Chaussidon}, E. and {Chaves-Montero}, J. and {Chen}, S. and {Chen}, X. and {Claybaugh}, T. and {Cole}, S. and {Cuceu}, A. and {Davis}, T.~M. and {Dawson}, K. and {de la Macorra}, A. and {de Mattia}, A. and {Deiosso}, N. and {Dey}, A. and {Dey}, B. and {Ding}, Z. and {Doel}, P. and {Edelstein}, J. and {Eftekharzadeh}, S. and {Eisenstein}, D.~J. and {Elliott}, A. and {Fagrelius}, P. and {Fanning}, K. and {Ferraro}, S. and {Ereza}, J. and {Findlay}, N. and {Flaugher}, B. and {Font-Ribera}, A. and {Forero-S{\'a}nchez}, D. and {Forero-Romero}, J.~E. and {Frenk}, C.~S. and {Garcia-Quintero}, C. and {Gazta{\~n}aga}, E. and {Gil-Mar{\'\i}n}, H. and {Gontcho}, S. Gontcho A and {Gonzalez-Morales}, A.~X. and {Gonzalez-Perez}, V. and {Gordon}, C. and {Green}, D. and {Gruen}, D. and {Gsponer}, R. and {Gutierrez}, G. and {Guy}, J. and {Hadzhiyska}, B. and {Hahn}, C. and {Hanif}, M.~M. S and {Herrera-Alcantar}, H.~K. and {Honscheid}, K. and {Howlett}, C. and {Huterer}, D. and {Ir{\v{s}}i{\v{c}}}, V. and {Ishak}, M. and {Juneau}, S. and {Kara{\c{c}}ayl{\i}}, N.~G. and {Kehoe}, R. and {Kent}, S. and {Kirkby}, D. and {Kremin}, A. and {Krolewski}, A. and {Lai}, Y. and {Lan}, T. -W. and {Landriau}, M. and {Lang}, D. and {Lasker}, J. and {Le Goff}, J.~M. and {Le Guillou}, L. and {Leauthaud}, A. and {Levi}, M.~E. and {Li}, T.~S. and {Linder}, E. and {Lodha}, K. and {Magneville}, C. and {Manera}, M. and {Margala}, D. and {Martini}, P. and {Maus}, M. and {McDonald}, P. and {Medina-Varela}, L. and {Meisner}, A. and {Mena-Fern{\'a}ndez}, J. and {Miquel}, R. and {Moon}, J. and {Moore}, S. and {Moustakas}, J. and {Mudur}, N. and {Mueller}, E. and {Mu{\~n}oz-Guti{\'e}rrez}, A. and {Myers}, A.~D. and {Nadathur}, S. and {Napolitano}, L. and {Neveux}, R. and {Newman}, J.~A. and {Nguyen}, N.~M. and {Nie}, J. and {Niz}, G. and {Noriega}, H.~E. and {Padmanabhan}, N. and {Paillas}, E. and {Palanque-Delabrouille}, N. and {Pan}, J. and {Penmetsa}, S. and {Percival}, W.~J. and {Pieri}, M. and {Pinon}, M. and {Poppett}, C. and {Porredon}, A. and {Prada}, F. and {P{\'e}rez-Fern{\'a}ndez}, A. and {P{\'e}rez-R{\`a}fols}, I. and {Rabinowitz}, D. and {Raichoor}, A. and {Ram{\'\i}rez-P{\'e}rez}, C. and {Ramirez-Solano}, S. and {Rashkovetskyi}, M. and {Rezaie}, M. and {Rich}, J. and {Rocher}, A. and {Rockosi}, C. and {Roe}, N.~A. and {Rosado-Marin}, A. and {Ross}, A.~J. and {Rossi}, G. and {Ruggeri}, R. and {Ruhlmann-Kleider}, V. and {Samushia}, L. and {Sanchez}, E. and {Saulder}, C. and {Schlafly}, E.~F. and {Schlegel}, D. and {Schubnell}, M. and {Seo}, H. and {Shafieloo}, A. and {Sharples}, R. and {Silber}, J. and {Slosar}, A. and {Smith}, A. and {Sprayberry}, D. and {Tan}, T. and {Tarl{\'e}}, G. and {Taylor}, P. and {Trusov}, S. and {Ure{\~n}a-L{\'o}pez}, L.~A. and {Vaisakh}, R. and {Valcin}, D. and {Valdes}, F. and {Vargas-Maga{\~n}a}, M. and {Verde}, L. and {Walther}, M. and {Wang}, B. and {Wang}, M.~S. and {Weaver}, B.~A. and {Weaverdyck}, N. and {Wechsler}, R.~H. and {Weinberg}, D.~H. and {White}, M. and {Yu}, J. and {Yu}, Y. and {Yuan}, S. and {Y{\`e}che}, C. and {Zaborowski}, E.~A. and {Zarrouk}, P. and {Zhang}, H. and {Zhao}, C. and {Zhao}, R. and {Zhou}, R. and {Zhuang}, T. and {Zou}, H.},
        title = "{DESI 2024 VI: Cosmological Constraints from the Measurements of Baryon Acoustic Oscillations}",
      journal = {arXiv e-prints},
         year = 2024,
        month = apr,
          eid = {arXiv:2404.03002},
        pages = {arXiv:2404.03002},
          doi = {10.48550/arXiv.2404.03002},
archivePrefix = {arXiv},
       eprint = {2404.03002},
 primaryClass = {astro-ph.CO},
       adsurl = {https://ui.adsabs.harvard.edu/abs/2024arXiv240403002D}
}

@ARTICLE{2009JCAP...08..020M,
   author = {{McDonald}, P. and {Roy}, A.},
    title = "{Clustering of dark matter tracers: generalizing bias for the coming era of precision LSS}",
  journal = {\jcap},
archivePrefix = "arXiv",
   eprint = {0902.0991},
 primaryClass = "astro-ph.CO",
     year = 2009,
    month = aug,
   volume = 8,
      eid = {020},
    pages = {20},
      doi = {10.1088/1475-7516/2009/08/020},
   adsurl = {http://adsabs.harvard.edu/abs/2009JCAP...08..020M}
}

@ARTICLE{Murgia:2018,
       author = {{Murgia}, Riccardo and {Ir{\v{s}}i{\v{c}}}, Vid and {Viel}, Matteo},
        title = "{Novel constraints on noncold, nonthermal dark matter from Lyman-{\ensuremath{\alpha}} forest data}",
      journal = {\prd},
         year = 2018,
        month = oct,
       volume = {98},
       number = {8},
          eid = {083540},
        pages = {083540},
          doi = {10.1103/PhysRevD.98.083540},
archivePrefix = {arXiv},
       eprint = {1806.08371},
 primaryClass = {astro-ph.CO},
       adsurl = {https://ui.adsabs.harvard.edu/abs/2018PhRvD..98h3540M}
}

@ARTICLE{Garzilli:2019,
       author = {{Garzilli}, Antonella and {Magalich}, Andrii and {Theuns}, Tom and {Frenk}, Carlos S. and {Weniger}, Christoph and {Ruchayskiy}, Oleg and {Boyarsky}, Alexey},
        title = "{The Lyman-{\ensuremath{\alpha}} forest as a diagnostic of the nature of the dark matter}",
      journal = {\mnras},
         year = 2019,
        month = nov,
       volume = {489},
       number = {3},
        pages = {3456-3471},
          doi = {10.1093/mnras/stz2188},
archivePrefix = {arXiv},
       eprint = {1809.06585},
 primaryClass = {astro-ph.CO},
       adsurl = {https://ui.adsabs.harvard.edu/abs/2019MNRAS.489.3456G}
}

@ARTICLE{Rogers:2022,
       author = {{Rogers}, Keir K. and {Dvorkin}, Cora and {Peiris}, Hiranya V.},
        title = "{Limits on the Light Dark Matter-Proton Cross Section from Cosmic Large-Scale Structure}",
      journal = {\prl},
         year = 2022,
        month = apr,
       volume = {128},
       number = {17},
          eid = {171301},
        pages = {171301},
          doi = {10.1103/PhysRevLett.128.171301},
archivePrefix = {arXiv},
       eprint = {2111.10386},
 primaryClass = {astro-ph.CO},
       adsurl = {https://ui.adsabs.harvard.edu/abs/2022PhRvL.128q1301R}
}

@article{McDonald:2001,
    author = "McDonald, Patrick",
    title = "{Toward a measurement of the cosmological geometry at Z 2: predicting lyman-alpha forest correlation in three dimensions, and the potential of future data sets}",
    eprint = "astro-ph/0108064",
    archivePrefix = "arXiv",
    doi = "10.1086/345945",
    journal = "Astrophys. J.",
    volume = "585",
    pages = "34--51",
    year = "2003"
}

@article{Planck:2018vyg,
    author = "Aghanim, N. and others",
    collaboration = "Planck",
    title = "{Planck 2018 results. VI. Cosmological parameters}",
    eprint = "1807.06209",
    archivePrefix = "arXiv",
    primaryClass = "astro-ph.CO",
    doi = "10.1051/0004-6361/201833910",
    journal = "Astron. Astrophys.",
    volume = "641",
    pages = "A6",
    year = "2020",
    note = "[Erratum: Astron.Astrophys. 652, C4 (2021)]"
}

@ARTICLE{Chabanier:2019,
       author = {{Chabanier}, Sol{\`e}ne and {Palanque-Delabrouille}, Nathalie and {Y{\`e}che}, Christophe and {Le Goff}, Jean-Marc and {Armengaud}, Eric and {Bautista}, Julian and {Blomqvist}, Michael and {Busca}, Nicolas and {Dawson}, Kyle and {Etourneau}, Thomas and {Font-Ribera}, Andreu and {Lee}, Youngbae and {du Mas des Bourboux}, H{\'e}lion and {Pieri}, Matthew and {Rich}, James and {Rossi}, Graziano and {Schneider}, Donald and {Slosar}, An{\v{z}}e},
        title = "{The one-dimensional power spectrum from the SDSS DR14 Ly{\ensuremath{\alpha}} forests}",
      journal = {\jcap},
         year = 2019,
        month = jul,
       volume = {2019},
       number = {7},
          eid = {017},
        pages = {017},
          doi = {10.1088/1475-7516/2019/07/017},
archivePrefix = {arXiv},
       eprint = {1812.03554},
 primaryClass = {astro-ph.CO},
       adsurl = {https://ui.adsabs.harvard.edu/abs/2019JCAP...07..017C}
}

@ARTICLE{2024MNRAS.tmp..176K,
       author = {{Kara{\c{c}}ayl{\i}}, Naim G{\"o}ksel and {Martini}, Paul and {Guy}, Julien and {Ravoux}, Corentin and {Karim}, Marie Lynn Abdul and {Armengaud}, Eric and {Walther}, Michael and {Aguilar}, J. and {Ahlen}, S. and {Bailey}, S. and {Bautista}, J. and {Beltran}, S.~F. and {Brooks}, D. and {Cabayol-Garcia}, L. and {Chabanier}, S. and {Chaussidon}, E. and {Chaves-Montero}, J. and {Dawson}, K. and {de la Cruz}, R. and {de la Macorra}, A. and {Doel}, P. and {Font-Ribera}, A. and {Forero-Romero}, J.~E. and {Gontcho}, S. Gontcho A. and {Gonzalez-Morales}, A.~X. and {Gordon}, C. and {Herrera-Alcantar}, H.~K. and {Honscheid}, K. and {Ir{\v{s}}i{\v{c}}}, V. and {Ishak}, M. and {Kehoe}, R. and {Kisner}, T. and {Kremin}, A. and {Landriau}, M. and {Le Guillou}, L. and {Levi}, M.~E. and {Luki{\'c}}, Z. and {Meisner}, A. and {Miquel}, R. and {Moustakas}, J. and {Mueller}, E. and {Mu{\~n}oz-Guti{\'e}rrez}, A. and {Napolitano}, L. and {Nie}, J. and {Niz}, G. and {Palanque-Delabrouille}, N. and {Percival}, W.~J. and {Pieri}, M. and {Poppett}, C. and {Prada}, F. and {P{\'e}rez-R{\`a}fols}, I. and {Ram{\'\i}rez-P{\'e}rez}, C. and {Rossi}, G. and {Sanchez}, E. and {Seo}, H. and {Sinigaglia}, F. and {Tan}, T. and {Tarl{\'e}}, G. and {Wang}, B. and {Weaver}, B.~A. and {Y{\'e}che}, C. and {Zhou}, Z.},
        title = "{Optimal 1D Ly{\ensuremath{\alpha}} Forest Power Spectrum Estimation - III. DESI early data}",
      journal = {\mnras},
         year = 2024,
        month = jan,
          doi = {10.1093/mnras/stae171},
archivePrefix = {arXiv},
       eprint = {2306.06316},
 primaryClass = {astro-ph.CO},
       adsurl = {https://ui.adsabs.harvard.edu/abs/2024MNRAS.tmp..176K}
}

@ARTICLE{2023MNRAS.526.5118R,
       author = {{Ravoux}, Corentin and {Abdul Karim}, Marie Lynn and {Armengaud}, Eric and {Walther}, Michael and {Kara{\c{c}}ayl{\i}}, Naim G{\"o}ksel and {Martini}, Paul and {Guy}, Julien and {Aguilar}, Jessica Nicole and {Ahlen}, Steven and {Bailey}, Stephen and {Bautista}, Julian and {Beltran}, Sergio Felipe and {Brooks}, David and {Cabayol-Garcia}, Laura and {Chabanier}, Sol{\`e}ne and {Chaussidon}, Edmond and {Chaves-Montero}, Jon{\'a}s and {Dawson}, Kyle and {de la Cruz}, Rodrigo and {de la Macorra}, Axel and {Doel}, Peter and {Fanning}, Kevin and {Font-Ribera}, Andreu and {Forero-Romero}, Jaime and {Gontcho A Gontcho}, Satya and {Gonzalez-Morales}, Alma X. and {Gordon}, Calum and {Herrera-Alcantar}, Hiram K. and {Honscheid}, Klaus and {Ir{\v{s}}i{\v{c}}}, Vid and {Ishak}, Mustapha and {Kehoe}, Robert and {Kisner}, Theodore and {Kremin}, Anthony and {Landriau}, Martin and {Le Guillou}, Laurent and {Levi}, Michael and {Luki{\'c}}, Zarija and {Magneville}, Christophe and {Meisner}, Aaron and {Miquel}, Ramon and {Moustakas}, John and {Mueller}, Eva-Maria and {Mu{\~n}oz-Guti{\'e}rrez}, Andrea and {Napolitano}, Lucas and {Nie}, Jundan and {Niz}, Gustavo and {Palanque-Delabrouille}, Nathalie and {Percival}, Will and {P{\'e}rez-R{\`a}fols}, Ignasi and {Pieri}, Matthew and {Poppett}, Claire and {Prada}, Francisco and {Ram{\'\i}rez P{\'e}rez}, C{\'e}sar and {Rossi}, Graziano and {Sanchez}, Eusebio and {Schlegel}, David and {Schubnell}, Michael and {Seo}, Hee-Jong and {Sinigaglia}, Francesco and {Tan}, Ting and {Tarl{\'e}}, Gregory and {Wang}, Ben and {Weaver}, Benjamin and {Y{\`e}che}, Christophe and {Zhou}, Zhimin},
        title = "{The Dark Energy Spectroscopic Instrument: one-dimensional power spectrum from first Ly {\ensuremath{\alpha}} forest samples with Fast Fourier Transform}",
      journal = {\mnras},
         year = 2023,
        month = dec,
       volume = {526},
       number = {4},
        pages = {5118-5140},
          doi = {10.1093/mnras/stad3008},
archivePrefix = {arXiv},
       eprint = {2306.06311},
 primaryClass = {astro-ph.CO},
       adsurl = {https://ui.adsabs.harvard.edu/abs/2023MNRAS.526.5118R}
}

@ARTICLE{Palanque2020,
       author = {{Palanque-Delabrouille}, Nathalie and {Y{\`e}che}, Christophe and {Sch{\"o}neberg}, Nils and {Lesgourgues}, Julien and {Walther}, Michael and {Chabanier}, Sol{\`e}ne and {Armengaud}, Eric},
        title = "{Hints, neutrino bounds, and WDM constraints from SDSS DR14 Lyman-{\ensuremath{\alpha}} and Planck full-survey data}",
      journal = {\jcap},
         year = 2020,
        month = apr,
       volume = {2020},
       number = {4},
          eid = {038},
        pages = {038},
          doi = {10.1088/1475-7516/2020/04/038},
archivePrefix = {arXiv},
       eprint = {1911.09073},
 primaryClass = {astro-ph.CO},
       adsurl = {https://ui.adsabs.harvard.edu/abs/2020JCAP...04..038P}
}

@article{Murgia:2019,
    author = "Murgia, Riccardo and Scelfo, Giulio and Viel, Matteo and Raccanelli, Alvise",
    title = "{Lyman-\ensuremath{\alpha} Forest Constraints on Primordial Black Holes as Dark Matter}",
    eprint = "1903.10509",
    archivePrefix = "arXiv",
    primaryClass = "astro-ph.CO",
    reportNumber = "CERN-TH-2019-029",
    doi = "10.1103/PhysRevLett.123.071102",
    journal = "Phys. Rev. Lett.",
    volume = "123",
    number = "7",
    pages = "071102",
    year = "2019"
}

@ARTICLE{Afshordi:2003,
       author = {{Afshordi}, N. and {McDonald}, P. and {Spergel}, D.~N.},
        title = "{Primordial Black Holes as Dark Matter: The Power Spectrum and Evaporation of Early Structures}",
      journal = {\apjl},
         year = 2003,
        month = sep,
       volume = {594},
       number = {2},
        pages = {L71-L74},
          doi = {10.1086/378763},
archivePrefix = {arXiv},
       eprint = {astro-ph/0302035},
 primaryClass = {astro-ph},
       adsurl = {https://ui.adsabs.harvard.edu/abs/2003ApJ...594L..71A}
}

@ARTICLE{Irsic:2020,
       author = {{Ir{\v{s}}i{\v{c}}}, Vid and {Xiao}, Huangyu and {McQuinn}, Matthew},
        title = "{Early structure formation constraints on the ultralight axion in the postinflation scenario}",
      journal = {\prd},
         year = 2020,
        month = jun,
       volume = {101},
       number = {12},
          eid = {123518},
        pages = {123518},
          doi = {10.1103/PhysRevD.101.123518},
archivePrefix = {arXiv},
       eprint = {1911.11150},
 primaryClass = {astro-ph.CO},
       adsurl = {https://ui.adsabs.harvard.edu/abs/2020PhRvD.101l3518I}
}

@ARTICLE{Villasenor:2023,
       author = {{Villasenor}, Bruno and {Robertson}, Brant and {Madau}, Piero and {Schneider}, Evan},
        title = "{New constraints on warm dark matter from the Lyman-{\ensuremath{\alpha}} forest power spectrum}",
      journal = {\prd},
         year = 2023,
        month = jul,
       volume = {108},
       number = {2},
          eid = {023502},
        pages = {023502},
          doi = {10.1103/PhysRevD.108.023502},
archivePrefix = {arXiv},
       eprint = {2209.14220},
 primaryClass = {astro-ph.CO},
       adsurl = {https://ui.adsabs.harvard.edu/abs/2023PhRvD.108b3502V}
}

@ARTICLE{Irsic:2023,
       author = {{Ir{\v{s}}i{\v{c}}}, Vid and {Viel}, Matteo and {Haehnelt}, Martin G. and {Bolton}, James S. and {Molaro}, Margherita and {Puchwein}, Ewald and {Boera}, Elisa and {Becker}, George D. and {Gaikwad}, Prakash and {Keating}, Laura C. and {Kulkarni}, Girish},
        title = "{Unveiling Dark Matter free-streaming at the smallest scales with high redshift Lyman-alpha forest}",
      journal = {arXiv e-prints},
         year = 2023,
        month = sep,
          eid = {arXiv:2309.04533},
        pages = {arXiv:2309.04533},
          doi = {10.48550/arXiv.2309.04533},
archivePrefix = {arXiv},
       eprint = {2309.04533},
 primaryClass = {astro-ph.CO},
       adsurl = {https://ui.adsabs.harvard.edu/abs/2023arXiv230904533I}
}

@ARTICLE{Villasenor:2022,
       author = {{Villasenor}, Bruno and {Robertson}, Brant and {Madau}, Piero and {Schneider}, Evan},
        title = "{Inferring the Thermal History of the Intergalactic Medium from the Properties of the Hydrogen and Helium Ly{\ensuremath{\alpha}} Forest}",
      journal = {\apj},
         year = 2022,
        month = jul,
       volume = {933},
       number = {1},
          eid = {59},
        pages = {59},
          doi = {10.3847/1538-4357/ac704e},
archivePrefix = {arXiv},
       eprint = {2111.00019},
 primaryClass = {astro-ph.CO},
       adsurl = {https://ui.adsabs.harvard.edu/abs/2022ApJ...933...59V}
}

@ARTICLE{2023PhRvL.131t1001G,
       author = {{Goldstein}, Samuel and {Hill}, J. Colin and {Ir{\v{s}}i{\v{c}}}, Vid and {Sherwin}, Blake D.},
        title = "{Canonical Hubble-Tension-Resolving Early Dark Energy Cosmologies Are Inconsistent with the Lyman-{\ensuremath{\alpha}} Forest}",
      journal = {\prl},
         year = 2023,
        month = nov,
       volume = {131},
       number = {20},
          eid = {201001},
        pages = {201001},
          doi = {10.1103/PhysRevLett.131.201001},
archivePrefix = {arXiv},
       eprint = {2303.00746},
 primaryClass = {astro-ph.CO},
       adsurl = {https://ui.adsabs.harvard.edu/abs/2023PhRvL.131t1001G}
}

@ARTICLE{Gaikwad:2021,
       author = {{Gaikwad}, Prakash and {Srianand}, Raghunathan and {Haehnelt}, Martin G. and {Choudhury}, Tirthankar Roy},
        title = "{A consistent and robust measurement of the thermal state of the IGM at 2 {\ensuremath{\leq}} z {\ensuremath{\leq}} 4 from a large sample of Ly {\ensuremath{\alpha}} forest spectra: evidence for late and rapid He II reionization}",
      journal = {\mnras},
         year = 2021,
        month = sep,
       volume = {506},
       number = {3},
        pages = {4389-4412},
          doi = {10.1093/mnras/stab2017},
archivePrefix = {arXiv},
       eprint = {2009.00016},
 primaryClass = {astro-ph.CO},
       adsurl = {https://ui.adsabs.harvard.edu/abs/2021MNRAS.506.4389G}
}

@ARTICLE{2008PhRvD..78h3519M,
   author = {{Matsubara}, T.},
    title = "{Nonlinear perturbation theory with halo bias and redshift-space distortions via the Lagrangian picture}",
  journal = {\prd},
archivePrefix = "arXiv",
   eprint = {0807.1733},
     year = 2008,
    month = oct,
   volume = 78,
   number = 8,
      eid = {083519},
    pages = {083519},
      doi = {10.1103/PhysRevD.78.083519},
   adsurl = {http://adsabs.harvard.edu/abs/2008PhRvD..78h3519M}
}

@ARTICLE{Font:2013,
       author = {{Font-Ribera}, Andreu and {Arnau}, Eduard and {Miralda-Escud{\'e}}, Jordi and {Rollinde}, Emmanuel and {Brinkmann}, J. and {Brownstein}, Joel R. and {Lee}, Khee-Gan and {Myers}, Adam D. and {Palanque-Delabrouille}, Nathalie and {P{\^a}ris}, Isabelle and {Petitjean}, Patrick and {Rich}, James and {Ross}, Nicholas P. and {Schneider}, Donald P. and {White}, Martin},
        title = "{The large-scale quasar-Lyman {\ensuremath{\alpha}} forest cross-correlation from BOSS}",
      journal = {\jcap},
         year = 2013,
        month = may,
       volume = {2013},
       number = {5},
          eid = {018},
        pages = {018},
          doi = {10.1088/1475-7516/2013/05/018},
archivePrefix = {arXiv},
       eprint = {1303.1937},
 primaryClass = {astro-ph.CO},
       adsurl = {https://ui.adsabs.harvard.edu/abs/2013JCAP...05..018F}
}

@ARTICLE{McDonald06,
       author = {{McDonald}, Patrick and {Seljak}, Uro{\v{s}} and {Burles}, Scott and
         {Schlegel}, David J. and {Weinberg}, David H. and {Cen}, Renyue and
         {Shih}, David and {Schaye}, Joop and {Schneider}, Donald P. and
         {Bahcall}, Neta A. and {Briggs}, John W. and {Brinkmann}, J. and
         {Brunner}, Robert J. and {Fukugita}, Masataka and {Gunn}, James E. and
         {Ivezi{\'c}}, {\v{Z}}eljko and {Kent}, Stephen and {Lupton}, Robert H. and
         {Vanden Berk}, Daniel E.},
        title = "{The Ly{\ensuremath{\alpha}} Forest Power Spectrum from the Sloan Digital Sky Survey}",
      journal = {\apjs},
         year = 2006,
        month = mar,
       volume = {163},
       number = {1},
        pages = {80-109},
          doi = {10.1086/444361},
archivePrefix = {arXiv},
       eprint = {astro-ph/0405013},
 primaryClass = {astro-ph},
       adsurl = {https://ui.adsabs.harvard.edu/abs/2006ApJS..163...80M}
}

@ARTICLE{1970Ap&SS...7....3S,
   author = {{Sunyaev}, R.~A. and {Zeldovich}, Y.~B.},
    title = "{Small-Scale Fluctuations of Relic Radiation}",
  journal = {\apss},
     year = 1970,
    month = apr,
   volume = 7,
    pages = {3-19},
      doi = {10.1007/BF00653471},
   adsurl = {http://adsabs.harvard.edu/abs/1970Ap%26SS...7....3S}
}

@ARTICLE{1970ApJ...162..815P,
   author = {{Peebles}, P.~J.~E. and {Yu}, J.~T.},
    title = "{Primeval Adiabatic Perturbation in an Expanding Universe}",
  journal = {\apj},
     year = 1970,
    month = dec,
   volume = 162,
    pages = {815},
      doi = {10.1086/150713},
   adsurl = {http://adsabs.harvard.edu/abs/1970ApJ...162..815P}
}

@ARTICLE{McDonald:2007,
       author = {{McDonald}, Patrick and {Eisenstein}, Daniel J.},
        title = "{Dark energy and curvature from a future baryonic acoustic oscillation survey using the Lyman-{\ensuremath{\alpha}} forest}",
      journal = {\prd},
         year = 2007,
        month = sep,
       volume = {76},
       number = {6},
          eid = {063009},
        pages = {063009},
          doi = {10.1103/PhysRevD.76.063009},
archivePrefix = {arXiv},
       eprint = {astro-ph/0607122},
 primaryClass = {astro-ph},
       adsurl = {https://ui.adsabs.harvard.edu/abs/2007PhRvD..76f3009M}
}

@ARTICLE{Zaldarriaga:2002,
       author = {{Zaldarriaga}, Matias},
        title = "{Searching for Fluctuations in the Intergalactic Medium Temperature Using the Ly{\ensuremath{\alpha}} Forest}",
      journal = {\apj},
         year = 2002,
        month = jan,
       volume = {564},
       number = {1},
        pages = {153-161},
          doi = {10.1086/324212},
archivePrefix = {arXiv},
       eprint = {astro-ph/0102205},
 primaryClass = {astro-ph},
       adsurl = {https://ui.adsabs.harvard.edu/abs/2002ApJ...564..153Z}
}

@ARTICLE{McQuinn:2016,
       author = {{McQuinn}, Matthew},
        title = "{The Evolution of the Intergalactic Medium}",
      journal = {\araa},
         year = 2016,
        month = sep,
       volume = {54},
        pages = {313-362},
          doi = {10.1146/annurev-astro-082214-122355},
archivePrefix = {arXiv},
       eprint = {1512.00086},
 primaryClass = {astro-ph.CO},
       adsurl = {https://ui.adsabs.harvard.edu/abs/2016ARA&A..54..313M}
}

@ARTICLE{Bolton:2008,
       author = {{Bolton}, J.~S. and {Viel}, M. and {Kim}, T. -S. and {Haehnelt}, M.~G. and {Carswell}, R.~F.},
        title = "{Possible evidence for an inverted temperature-density relation in the intergalactic medium from the flux distribution of the Ly{\ensuremath{\alpha}} forest}",
      journal = {\mnras},
         year = 2008,
        month = may,
       volume = {386},
       number = {2},
        pages = {1131-1144},
          doi = {10.1111/j.1365-2966.2008.13114.x},
archivePrefix = {arXiv},
       eprint = {0711.2064},
 primaryClass = {astro-ph},
       adsurl = {https://ui.adsabs.harvard.edu/abs/2008MNRAS.386.1131B}
}

@ARTICLE{Walther:2019,
       author = {{Walther}, Michael and {O{\~n}orbe}, Jose and {Hennawi}, Joseph F. and {Luki{\'c}}, Zarija},
        title = "{New Constraints on IGM Thermal Evolution from the Ly{\ensuremath{\alpha}} Forest Power Spectrum}",
      journal = {\apj},
         year = 2019,
        month = feb,
       volume = {872},
       number = {1},
          eid = {13},
        pages = {13},
          doi = {10.3847/1538-4357/aafad1},
archivePrefix = {arXiv},
       eprint = {1808.04367},
 primaryClass = {astro-ph.CO},
       adsurl = {https://ui.adsabs.harvard.edu/abs/2019ApJ...872...13W}
}

@ARTICLE{Viel:2006,
       author = {{Viel}, Matteo and {Lesgourgues}, Julien and {Haehnelt}, Martin G. and {Matarrese}, Sabino and {Riotto}, Antonio},
        title = "{Can Sterile Neutrinos Be Ruled Out as Warm Dark Matter Candidates?}",
      journal = {\prl},
         year = 2006,
        month = aug,
       volume = {97},
       number = {7},
          eid = {071301},
        pages = {071301},
          doi = {10.1103/PhysRevLett.97.071301},
archivePrefix = {arXiv},
       eprint = {astro-ph/0605706},
 primaryClass = {astro-ph},
       adsurl = {https://ui.adsabs.harvard.edu/abs/2006PhRvL..97g1301V}
}

@ARTICLE{Meiksin:2009,
       author = {{Meiksin}, Avery A.},
        title = "{The physics of the intergalactic medium}",
      journal = {Reviews of Modern Physics},
         year = 2009,
        month = oct,
       volume = {81},
       number = {4},
        pages = {1405-1469},
          doi = {10.1103/RevModPhys.81.1405},
archivePrefix = {arXiv},
       eprint = {0711.3358},
 primaryClass = {astro-ph},
       adsurl = {https://ui.adsabs.harvard.edu/abs/2009RvMP...81.1405M}
}

@ARTICLE{Busca:2013,
       author = {{Busca}, N.~G. and {Delubac}, T. and {Rich}, J. and {Bailey}, S. and {Font-Ribera}, A. and {Kirkby}, D. and {Le Goff}, J. -M. and {Pieri}, M.~M. and {Slosar}, A. and {Aubourg}, {\'E}. and {Bautista}, J.~E. and {Bizyaev}, D. and {Blomqvist}, M. and {Bolton}, A.~S. and {Bovy}, J. and {Brewington}, H. and {Borde}, A. and {Brinkmann}, J. and {Carithers}, B. and {Croft}, R.~A.~C. and {Dawson}, K.~S. and {Ebelke}, G. and {Eisenstein}, D.~J. and {Hamilton}, J. -C. and {Ho}, S. and {Hogg}, D.~W. and {Honscheid}, K. and {Lee}, K. -G. and {Lundgren}, B. and {Malanushenko}, E. and {Malanushenko}, V. and {Margala}, D. and {Maraston}, C. and {Mehta}, K. and {Miralda-Escud{\'e}}, J. and {Myers}, A.~D. and {Nichol}, R.~C. and {Noterdaeme}, P. and {Olmstead}, M.~D. and {Oravetz}, D. and {Palanque-Delabrouille}, N. and {Pan}, K. and {P{\^a}ris}, I. and {Percival}, W.~J. and {Petitjean}, P. and {Roe}, N.~A. and {Rollinde}, E. and {Ross}, N.~P. and {Rossi}, G. and {Schlegel}, D.~J. and {Schneider}, D.~P. and {Shelden}, A. and {Sheldon}, E.~S. and {Simmons}, A. and {Snedden}, S. and {Tinker}, J.~L. and {Viel}, M. and {Weaver}, B.~A. and {Weinberg}, D.~H. and {White}, M. and {Y{\`e}che}, C. and {York}, D.~G.},
        title = "{Baryon acoustic oscillations in the Ly{\ensuremath{\alpha}} forest of BOSS quasars}",
      journal = {\aap},
         year = 2013,
        month = apr,
       volume = {552},
          eid = {A96},
        pages = {A96},
          doi = {10.1051/0004-6361/201220724},
archivePrefix = {arXiv},
       eprint = {1211.2616},
 primaryClass = {astro-ph.CO},
       adsurl = {https://ui.adsabs.harvard.edu/abs/2013A&A...552A..96B}
}

@ARTICLE{Cuceu:2021,
       author = {{Cuceu}, Andrei and {Font-Ribera}, Andreu and {Joachimi}, Benjamin and {Nadathur}, Seshadri},
        title = "{Cosmology beyond BAO from the 3D distribution of the Lyman-{\ensuremath{\alpha}} forest}",
      journal = {\mnras},
         year = 2021,
        month = oct,
       volume = {506},
       number = {4},
        pages = {5439-5450},
          doi = {10.1093/mnras/stab1999},
archivePrefix = {arXiv},
       eprint = {2103.14075},
 primaryClass = {astro-ph.CO},
       adsurl = {https://ui.adsabs.harvard.edu/abs/2021MNRAS.506.5439C}
}

@ARTICLE{Wilson:2022,
       author = {{Wilson}, Bayu and {Ir{\v{s}}i{\v{c}}}, Vid and {McQuinn}, Matthew},
        title = "{A measurement of the Ly {\ensuremath{\beta}} forest power spectrum and its cross with the Ly {\ensuremath{\alpha}} forest in X-Shooter XQ-100}",
      journal = {\mnras},
         year = 2022,
        month = jan,
       volume = {509},
       number = {2},
        pages = {2423-2442},
          doi = {10.1093/mnras/stab3017},
archivePrefix = {arXiv},
       eprint = {2106.04837},
 primaryClass = {astro-ph.CO},
       adsurl = {https://ui.adsabs.harvard.edu/abs/2022MNRAS.509.2423W}
}

@ARTICLE{Irsic17,
   author = {{Ir{\v s}i{\v c}}, V. and {Viel}, M. and {Haehnelt}, M.~G. and 
	{Bolton}, J.~S. and {Cristiani}, S. and {Becker}, G.~D. and 
	{D'Odorico}, V. and {Cupani}, G. and {Kim}, T.-S. and {Berg}, T.~A.~M. and 
	{L{\'o}pez}, S. and {Ellison}, S. and {Christensen}, L. and 
	{Denny}, K.~D. and {Worseck}, G.},
    title = "{New Constraints on the free-streaming of warm dark matter from intermediate and small scale Lyman-$\alpha$ forest data}",
  journal = {ArXiv e-prints},
archivePrefix = "arXiv",
   eprint = {1702.01764},
     year = 2017,
    month = feb,
   adsurl = {http://adsabs.harvard.edu/abs/2017arXiv170201764I}
}

@ARTICLE{Planck,
   author = {{Planck Collaboration} and {Ade}, P.~A.~R. and {Aghanim}, N. and 
	{Arnaud}, M. and {Ashdown}, M. and {Aumont}, J. and {Baccigalupi}, C. and 
	{Banday}, A.~J. and {Barreiro}, R.~B. and {Bartlett}, J.~G. and et al.},
    title = "{Planck 2015 results. XIII. Cosmological parameters}",
  journal = {\aap},
archivePrefix = "arXiv",
   eprint = {1502.01589},
     year = 2016,
    month = sep,
   volume = 594,
      eid = {A13},
    pages = {A13},
      doi = {10.1051/0004-6361/201525830},
   adsurl = {http://adsabs.harvard.edu/abs/2016A%26A...594A..13P}
}

@ARTICLE{PYB13,
       author = {{Palanque-Delabrouille}, Nathalie and {Y{\`e}che}, Christophe and
         {Borde}, Arnaud and {Le Goff}, Jean-Marc and {Rossi}, Graziano and
         {Viel}, Matteo and {Aubourg}, {\'E}ric and {Bailey}, Stephen and
         {Bautista}, Julian and {Blomqvist}, Michael and {Bolton}, Adam and
         {Bolton}, James S. and {Busca}, Nicol{\'a}s G. and {Carithers}, Bill and
         {Croft}, Rupert A.~C. and {Dawson}, Kyle S. and
         {Delubac}, Timoth{\'e}e and {Font-Ribera}, Andreu and {Ho}, Shirley and
         {Kirkby}, David and {Lee}, Khee-Gan and {Margala}, Daniel and
         {Miralda-Escud{\'e}}, Jordi and {Muna}, Demitri and {Myers}, Adam D. and
         {Noterdaeme}, Pasquier and {P{\^a}ris}, Isabelle and
         {Petitjean}, Patrick and {Pieri}, Matthew M. and {Rich}, James and
         {Rollinde}, Emmanuel and {Ross}, Nicholas P. and {Schlegel}, David J. and
         {Schneider}, Donald P. and {Slosar}, An{\v{z}}e and
         {Weinberg}, David H.},
        title = "{The one-dimensional Ly{\ensuremath{\alpha}} forest power spectrum from BOSS}",
      journal = {\aap},
         year = 2013,
        month = nov,
       volume = {559},
          eid = {A85},
        pages = {A85},
          doi = {10.1051/0004-6361/201322130},
archivePrefix = {arXiv},
       eprint = {1306.5896},
 primaryClass = {astro-ph.CO},
       adsurl = {https://ui.adsabs.harvard.edu/abs/2013A&A...559A..85P}
}

@ARTICLE{Alam:2021,
       author = {{Alam}, Shadab and {Aubert}, Marie and {Avila}, Santiago and {Balland}, Christophe and {Bautista}, Julian E. and {Bershady}, Matthew A. and {Bizyaev}, Dmitry and {Blanton}, Michael R. and {Bolton}, Adam S. and {Bovy}, Jo and {Brinkmann}, Jonathan and {Brownstein}, Joel R. and {Burtin}, Etienne and {Chabanier}, Sol{\`e}ne and {Chapman}, Michael J. and {Choi}, Peter Doohyun and {Chuang}, Chia-Hsun and {Comparat}, Johan and {Cousinou}, Marie-Claude and {Cuceu}, Andrei and {Dawson}, Kyle S. and {de la Torre}, Sylvain and {de Mattia}, Arnaud and {Agathe}, Victoria de Sainte and {des Bourboux}, H{\'e}lion du Mas and {Escoffier}, Stephanie and {Etourneau}, Thomas and {Farr}, James and {Font-Ribera}, Andreu and {Frinchaboy}, Peter M. and {Fromenteau}, Sebastien and {Gil-Mar{\'\i}n}, H{\'e}ctor and {Le Goff}, Jean-Marc and {Gonzalez-Morales}, Alma X. and {Gonzalez-Perez}, Violeta and {Grabowski}, Kathleen and {Guy}, Julien and {Hawken}, Adam J. and {Hou}, Jiamin and {Kong}, Hui and {Parker}, James and {Klaene}, Mark and {Kneib}, Jean-Paul and {Lin}, Sicheng and {Long}, Daniel and {Lyke}, Brad W. and {de la Macorra}, Axel and {Martini}, Paul and {Masters}, Karen and {Mohammad}, Faizan G. and {Moon}, Jeongin and {Mueller}, Eva-Maria and {Mu{\~n}oz-Guti{\'e}rrez}, Andrea and {Myers}, Adam D. and {Nadathur}, Seshadri and {Neveux}, Richard and {Newman}, Jeffrey A. and {Noterdaeme}, Pasquier and {Oravetz}, Audrey and {Oravetz}, Daniel and {Palanque-Delabrouille}, Nathalie and {Pan}, Kaike and {Paviot}, Romain and {Percival}, Will J. and {P{\'e}rez-R{\`a}fols}, Ignasi and {Petitjean}, Patrick and {Pieri}, Matthew M. and {Prakash}, Abhishek and {Raichoor}, Anand and {Ravoux}, Corentin and {Rezaie}, Mehdi and {Rich}, James and {Ross}, Ashley J. and {Rossi}, Graziano and {Ruggeri}, Rossana and {Ruhlmann-Kleider}, Vanina and {S{\'a}nchez}, Ariel G. and {S{\'a}nchez}, F. Javier and {S{\'a}nchez-Gallego}, Jos{\'e} R. and {Sayres}, Conor and {Schneider}, Donald P. and {Seo}, Hee-Jong and {Shafieloo}, Arman and {Slosar}, An{\v{z}}e and {Smith}, Alex and {Stermer}, Julianna and {Tamone}, Amelie and {Tinker}, Jeremy L. and {Tojeiro}, Rita and {Vargas-Maga{\~n}a}, Mariana and {Variu}, Andrei and {Wang}, Yuting and {Weaver}, Benjamin A. and {Weijmans}, Anne-Marie and {Y{\`e}che}, Christophe and {Zarrouk}, Pauline and {Zhao}, Cheng and {Zhao}, Gong-Bo and {Zheng}, Zheng},
        title = "{Completed SDSS-IV extended Baryon Oscillation Spectroscopic Survey: Cosmological implications from two decades of spectroscopic surveys at the Apache Point Observatory}",
      journal = {\prd},
         year = 2021,
        month = apr,
       volume = {103},
       number = {8},
          eid = {083533},
        pages = {083533},
          doi = {10.1103/PhysRevD.103.083533},
archivePrefix = {arXiv},
       eprint = {2007.08991},
 primaryClass = {astro-ph.CO},
       adsurl = {https://ui.adsabs.harvard.edu/abs/2021PhRvD.103h3533A}
}

@ARTICLE{Slosar2013,
       author = {{Slosar}, An{\v{z}}e and {Ir{\v{s}}i{\v{c}}}, Vid and {Kirkby}, David and {Bailey}, Stephen and {Busca}, Nicol{\'a}s G. and {Delubac}, Timoth{\'e}e and {Rich}, James and {Aubourg}, {\'E}ric and {Bautista}, Julian E. and {Bhardwaj}, Vaishali and {Blomqvist}, Michael and {Bolton}, Adam S. and {Bovy}, Jo and {Brownstein}, Joel and {Carithers}, Bill and {Croft}, Rupert A.~C. and {Dawson}, Kyle S. and {Font-Ribera}, Andreu and {Le Goff}, J. -M. and {Ho}, Shirley and {Honscheid}, Klaus and {Lee}, Khee-Gan and {Margala}, Daniel and {McDonald}, Patrick and {Medolin}, Bumbarija and {Miralda-Escud{\'e}}, Jordi and {Myers}, Adam D. and {Nichol}, Robert C. and {Noterdaeme}, Pasquier and {Palanque-Delabrouille}, Nathalie and {P{\^a}ris}, Isabelle and {Petitjean}, Patrick and {Pieri}, Matthew M. and {Pi{\v{s}}kur}, Yodovina and {Roe}, Natalie A. and {Ross}, Nicholas P. and {Rossi}, Graziano and {Schlegel}, David J. and {Schneider}, Donald P. and {Suzuki}, Nao and {Sheldon}, Erin S. and {Seljak}, Uro{\v{s}} and {Viel}, Matteo and {Weinberg}, David H. and {Y{\`e}che}, Christophe},
        title = "{Measurement of baryon acoustic oscillations in the Lyman-{\ensuremath{\alpha}} forest fluctuations in BOSS data release 9}",
      journal = {\jcap},
         year = 2013,
        month = apr,
       volume = {2013},
       number = {4},
          eid = {026},
        pages = {026},
          doi = {10.1088/1475-7516/2013/04/026},
archivePrefix = {arXiv},
       eprint = {1301.3459},
 primaryClass = {astro-ph.CO},
       adsurl = {https://ui.adsabs.harvard.edu/abs/2013JCAP...04..026S}
}

@ARTICLE{Font-Ribera:2018,
       author = {{Font-Ribera}, Andreu and {McDonald}, Patrick and {Slosar}, An{\v{z}}e},
        title = "{How to estimate the 3D power spectrum of the Lyman-{\ensuremath{\alpha}} forest}",
      journal = {\jcap},
         year = 2018,
        month = jan,
       volume = {2018},
       number = {1},
          eid = {003},
        pages = {003},
          doi = {10.1088/1475-7516/2018/01/003},
archivePrefix = {arXiv},
       eprint = {1710.11036},
 primaryClass = {astro-ph.CO},
       adsurl = {https://ui.adsabs.harvard.edu/abs/2018JCAP...01..003F}
}

@ARTICLE{dMdB:2020,
       author = {{du Mas des Bourboux}, H{\'e}lion and {Rich}, James and {Font-Ribera}, Andreu and {de Sainte Agathe}, Victoria and {Farr}, James and {Etourneau}, Thomas and {Le Goff}, Jean-Marc and {Cuceu}, Andrei and {Balland}, Christophe and {Bautista}, Julian E. and {Blomqvist}, Michael and {Brinkmann}, Jonathan and {Brownstein}, Joel R. and {Chabanier}, Sol{\`e}ne and {Chaussidon}, Edmond and {Dawson}, Kyle and {Gonz{\'a}lez-Morales}, Alma X. and {Guy}, Julien and {Lyke}, Brad W. and {de la Macorra}, Axel and {Mueller}, Eva-Maria and {Myers}, Adam D. and {Nitschelm}, Christian and {Mu{\~n}oz Guti{\'e}rrez}, Andrea and {Palanque-Delabrouille}, Nathalie and {Parker}, James and {Percival}, Will J. and {P{\'e}rez-R{\`a}fols}, Ignasi and {Petitjean}, Patrick and {Pieri}, Matthew M. and {Ravoux}, Corentin and {Rossi}, Graziano and {Schneider}, Donald P. and {Seo}, Hee-Jong and {Slosar}, An{\v{z}}e and {Stermer}, Julianna and {Vivek}, M. and {Y{\`e}che}, Christophe and {Youles}, Samantha},
        title = "{The Completed SDSS-IV Extended Baryon Oscillation Spectroscopic Survey: Baryon Acoustic Oscillations with Ly{\ensuremath{\alpha}} Forests}",
      journal = {\apj},
         year = 2020,
        month = oct,
       volume = {901},
       number = {2},
          eid = {153},
        pages = {153},
          doi = {10.3847/1538-4357/abb085},
archivePrefix = {arXiv},
       eprint = {2007.08995},
 primaryClass = {astro-ph.CO},
       adsurl = {https://ui.adsabs.harvard.edu/abs/2020ApJ...901..153D}
}

@ARTICLE{Seljak:2005,
       author = {{Seljak}, Uro{\v{s}} and {Makarov}, Alexey and {McDonald}, Patrick and {Anderson}, Scott F. and {Bahcall}, Neta A. and {Brinkmann}, J. and {Burles}, Scott and {Cen}, Renyue and {Doi}, Mamoru and {Gunn}, James E. and {Ivezi{\'c}}, {\v{Z}}eljko and {Kent}, Stephen and {Loveday}, Jon and {Lupton}, Robert H. and {Munn}, Jeffrey A. and {Nichol}, Robert C. and {Ostriker}, Jeremiah P. and {Schlegel}, David J. and {Schneider}, Donald P. and {Tegmark}, Max and {Berk}, Daniel E. and {Weinberg}, David H. and {York}, Donald G.},
        title = "{Cosmological parameter analysis including SDSS Ly{\ensuremath{\alpha}} forest and galaxy bias: Constraints on the primordial spectrum of fluctuations, neutrino mass, and dark energy}",
      journal = {\prd},
         year = 2005,
        month = may,
       volume = {71},
       number = {10},
          eid = {103515},
        pages = {103515},
          doi = {10.1103/PhysRevD.71.103515},
archivePrefix = {arXiv},
       eprint = {astro-ph/0407372},
 primaryClass = {astro-ph},
       adsurl = {https://ui.adsabs.harvard.edu/abs/2005PhRvD..71j3515S}
}

@ARTICLE{Viel:2010,
       author = {{Viel}, Matteo and {Haehnelt}, Martin G. and {Springel}, Volker},
        title = "{The effect of neutrinos on the matter distribution as probed by the intergalactic medium}",
      journal = {\jcap},
         year = 2010,
        month = jun,
       volume = {2010},
       number = {6},
          eid = {015},
        pages = {015},
          doi = {10.1088/1475-7516/2010/06/015},
archivePrefix = {arXiv},
       eprint = {1003.2422},
 primaryClass = {astro-ph.CO},
       adsurl = {https://ui.adsabs.harvard.edu/abs/2010JCAP...06..015V}
}

@ARTICLE{Baur:2016,
       author = {{Baur}, Julien and {Palanque-Delabrouille}, Nathalie and {Y{\`e}che}, Christophe and {Magneville}, Christophe and {Viel}, Matteo},
        title = "{Lyman-alpha forests cool warm dark matter}",
      journal = {\jcap},
         year = 2016,
        month = aug,
       volume = {2016},
       number = {8},
          eid = {012},
        pages = {012},
          doi = {10.1088/1475-7516/2016/08/012},
archivePrefix = {arXiv},
       eprint = {1512.01981},
 primaryClass = {astro-ph.CO},
       adsurl = {https://ui.adsabs.harvard.edu/abs/2016JCAP...08..012B}
}

@ARTICLE{Viel:2013,
       author = {{Viel}, Matteo and {Becker}, George D. and {Bolton}, James S. and {Haehnelt}, Martin G.},
        title = "{Warm dark matter as a solution to the small scale crisis: New constraints from high redshift Lyman-{\ensuremath{\alpha}} forest data}",
      journal = {\prd},
         year = 2013,
        month = aug,
       volume = {88},
       number = {4},
          eid = {043502},
        pages = {043502},
          doi = {10.1103/PhysRevD.88.043502},
archivePrefix = {arXiv},
       eprint = {1306.2314},
 primaryClass = {astro-ph.CO},
       adsurl = {https://ui.adsabs.harvard.edu/abs/2013PhRvD..88d3502V}
}

@article{Pedersen:2020,
    author = "Pedersen, Christian and Font-Ribera, Andreu and Rogers, Keir K. and McDonald, Patrick and Peiris, Hiranya V. and Pontzen, Andrew and Slosar, An\v{z}e",
    title = "{An emulator for the Lyman-$\alpha$ forest in beyond-$\Lambda$CDM cosmologies}",
    eprint = "2011.15127",
    archivePrefix = "arXiv",
    primaryClass = "astro-ph.CO",
    doi = "10.1088/1475-7516/2021/05/033",
    journal = "JCAP",
    volume = "05",
    pages = "033",
    year = "2021"
}

@ARTICLE{Boera:2019,
       author = {{Boera}, Elisa and {Becker}, George D. and {Bolton}, James S. and {Nasir}, Fahad},
        title = "{Revealing Reionization with the Thermal History of the Intergalactic Medium: New Constraints from the Ly{\ensuremath{\alpha}} Flux Power Spectrum}",
      journal = {\apj},
         year = 2019,
        month = feb,
       volume = {872},
       number = {1},
          eid = {101},
        pages = {101},
          doi = {10.3847/1538-4357/aafee4},
archivePrefix = {arXiv},
       eprint = {1809.06980},
 primaryClass = {astro-ph.CO},
       adsurl = {https://ui.adsabs.harvard.edu/abs/2019ApJ...872..101B}
}

@ARTICLE{Gaikwad:2019,
       author = {{Gaikwad}, Prakash and {Srianand}, Raghunathan and {Khaire}, Vikram and {Choudhury}, Tirthankar Roy},
        title = "{Effect of non-equilibrium ionization on derived physical conditions of the high-z intergalactic medium}",
      journal = {\mnras},
         year = 2019,
        month = dec,
       volume = {490},
       number = {2},
        pages = {1588-1604},
          doi = {10.1093/mnras/stz2692},
archivePrefix = {arXiv},
       eprint = {1812.01016},
 primaryClass = {astro-ph.CO},
       adsurl = {https://ui.adsabs.harvard.edu/abs/2019MNRAS.490.1588G}
}

@article{Cuceu:2025nvl,
    author = "Cuceu, Andrei and others",
    title = "{DESI DR1 Ly$\alpha$ forest: 3D full-shape analysis and cosmological constraints}",
    eprint = "2509.15308",
    archivePrefix = "arXiv",
    primaryClass = "astro-ph.CO",
    reportNumber = "FERMILAB-PUB-25-0693-PPD",
    month = "9",
    year = "2025"
}

\end{document}